\documentclass{egpubl}
\usepackage{pg2025s}
\usepackage{kotex}
\usepackage{amsmath}
\usepackage{fancyhdr}
\usepackage{subfigure}
\usepackage{amssymb}
\usepackage{xcolor} 
\usepackage{enumitem}
\usepackage{booktabs}

\usepackage[T1]{fontenc}
\usepackage{dfadobe}  
\usepackage{cuted}

\usepackage{cite}

\BibtexOrBiblatex
\electronicVersion
\PrintedOrElectronic

\WsConferencePaper

\ifpdf \usepackage[pdftex]{graphicx} \pdfcompresslevel=9
\else \usepackage[dvips]{graphicx} \fi

\usepackage{egweblnk}

\definecolor{crr}{rgb}{0,0,0}

\definecolor{cro}{rgb}{0,0,0}

\definecolor{ys}{rgb}{0,0,0}

\definecolor{rev}{rgb}{0,0,0}

\definecolor{chk}{rgb}{0,0,0}

  \title[B2F: End-to-End Body-to-Face Motion Generation with Style Reference]{B2F: End-to-End Body-to-Face Motion Generation with Style Reference}

\author[B. Jang, E. Jung \& Y. Lee]
{\parbox{\textwidth}{\centering
Bokyung Jang\orcid{0009-0001-4800-8576},
Eunho Jung\orcid{0009-0005-6652-5189},
and Yoonsang Lee\thanks{Corresponding author.}\orcid{0000-0002-0579-5987}
                    }
       \\ 
{\parbox{\textwidth}{\centering Hanyang University, Department of Computer Science, South Korea
       }
}
}

\makeatletter
\def\ps@titlepage{%
  \def\@oddhead{}%
  \def\@evenhead{}%
  \def\@oddfoot{}%
  \def\@evenfoot{}%
}
\def\@oddhead{}
\def\@evenhead{}
\def\@oddfoot{}
\def\@evenfoot{}
\def\@volume{}
\def\@journal{}
\def\@Copyright{}
\makeatother

\begin{document}

\teaser{
  \centering
  \includegraphics[trim=0 0 0 0, clip, width=0.9\textwidth]{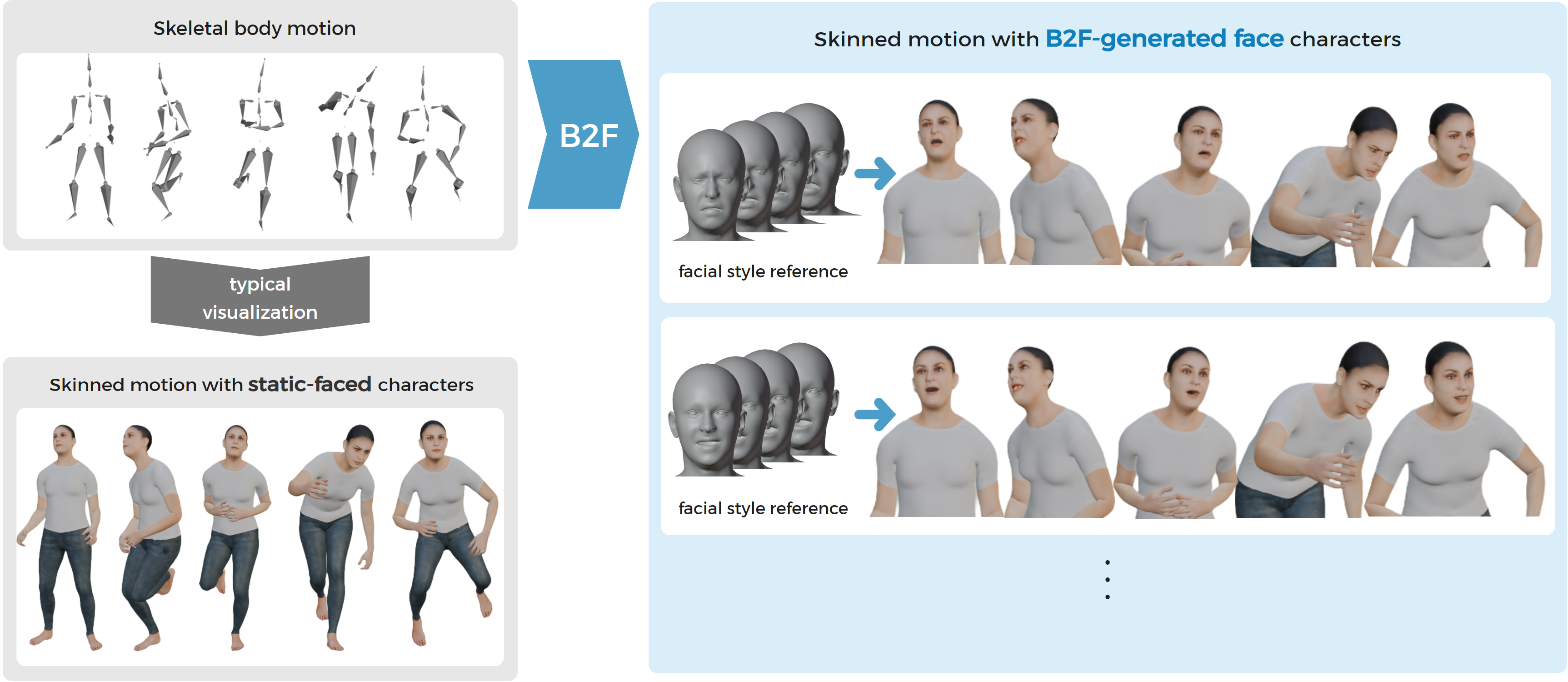}
  \caption{
  \textcolor{rev}{
    B2F generates facial expressions that align with body motions, enhancing the cohesiveness of character animation.
    Conditioned on a style reference, it produces diverse and expressive facial motions.}
    }
  }

\maketitle

\begin{abstract}
Human motion naturally integrates body movements and facial expressions, forming a unified perception.
If a virtual character’s facial expression does not align well with its body movements, it may weaken the perception of the character as a cohesive whole.
Motivated by this, we propose B2F, a model that generates facial motions aligned with body movements.
B2F takes a facial style reference as input, generating facial animations that reflect the provided style while maintaining consistency with the associated body motion.
To achieve this, B2F learns a disentangled representation of content and style, using alignment and consistency-based objectives.
We represent style using discrete latent codes learned via the Gumbel-Softmax trick, enabling diverse expression generation with a structured latent representation.
B2F outputs facial motion in the FLAME format, making it compatible with SMPL-X characters, and supports ARKit-style avatars through a dedicated conversion module.
Our evaluations show that B2F generates expressive and engaging facial animations that synchronize with body movements and style intent, while mitigating perceptual dissonance from mismatched cues, and generalizing across diverse characters and styles.

\printccsdesc  
\end{abstract}

\section{Introduction}
When a person performs an action, it is not just about moving their body; changes in facial expressions often accompany the motion.
\textcolor{ys}{Psychological and neuroscientific studies have consistently shown that humans perceive faces and bodies as an integrated whole-person representation rather than as independent modalities \cite{hu_integrating_2020}, reflecting a fundamental cognitive mechanism.}
If facial expressions are absent or misaligned with body movement, this integration may break down.
This raises the question: if a character's facial expressions do not align with its body motions, could it hinder the observer's perception and understanding?

Motivated by this question, we propose B2F, a method that generates facial motions aligned with given body motions, addressing the mismatch between body actions and facial expressions in animated characters.
\textcolor{ys}{In doing so, we define a new problem setting: body-to-face motion generation with style control. To the best of our knowledge, our work is the first to formulate this problem and propose a learning-based solution.}
Given that the same body motion can be paired with various facial styles, we assume that facial motion can be decomposed into two factors:
(1) content, which reflects the context-dependent expressions that naturally accompany body movements, and
(2) style, which captures the character’s expressive tendency or emotional tone.
For example, in the motion of lifting a heavy object, a strained expression during the lift followed by a relaxed face after placing it down reflects the content, while whether the character smiles or frowns during the process represents the style.
To better capture clear distinctions between facial styles, we adopt a discrete latent variable structure for style representation, learned using the Gumbel-Softmax trick 
\cite{jang2017categorical}.

Based on this decomposition, B2F is trained to enable controllable and consistent facial motion generation by disentangling content and style while preserving their respective characteristics in the output.
Specifically, we introduce an alignment loss that encourages the content representations extracted from body and facial motions to be similar when they come from the same segment, promoting a shared understanding of motion context.
In addition, we use consistency and cross-consistency losses to enforce that style and content information are both preserved in their respective embeddings, while remaining disentangled from each other.
These losses together help the model generate facial motions that are not only synchronized with the body motion but also carry the desired expressive tone.

\textcolor{rev}{
The B2F model consists of multiple encoders and a generator, all trained end-to-end.
It generates facial animations in the FLAME format~\cite{Li17}, making it inherently compatible with the SMPL-X human model~\cite{pavlakos2019expressive}.
Additionally, we provide a FLAME-to-ARKit converter that enables deployment to a wide range of humanoid characters supporting ARKit blendshapes, regardless of their mesh structure or skeleton configuration.
}

\textcolor{ys}{We validate the effectiveness of B2F through qualitative, quantitative, and perceptual evaluations. In particular, to empirically support our problem formulation, we conducted perceptual studies comparing animations with aligned facial motions against static or intentionally misaligned ones for the same body movements. Participants generally preferred the aligned animations over misaligned ones across expressiveness, engagement, and realism, confirming that semantic misalignment between body and facial motions can significantly harm user perception. Together, these results demonstrate that B2F generates expressive and engaging character animation by producing facial motions that align with both body movements and stylistic intent, while generalizing across diverse inputs and mitigating perceptual dissonance.}

\section{Related Work}
\paragraph*{Integrated Perception of Human Face and Body}

The perception of a person as a unified entity has been a key topic in psychology and neuroscience, with studies exploring how the interaction between facial and body cues influences our understanding of others.
Willis et al. \shortcite{willis_judging_2011} demonstrated that the interplay between facial and body expressions affects both approachability judgments and expression categorization.
Kret et al. \shortcite{kret_perception_2013} showed that emotional congruence between facial and body expressions enhances emotion recognition accuracy.
Interestingly, they found that even when participants were instructed to focus on body posture, they continued to pay significant attention to facial expressions.
Dynamic cues, particularly movement, have been identified as crucial in the integrated perception.
Pilz et al. \shortcite{pilz_walk_2011} found that visualized body motion can accelerate facial recognition,
while Simhi et al. \shortcite{simhi_contribution_2016} revealed that body motion contributes to person recognition beyond what is possible with facial information alone.
Neuroscientific studies using fMRI have explored the neural mechanisms supporting this integrated perception.
Cox et al. \shortcite{cox_contextually_2004} found that the fusiform face area (FFA), known for facial processing, is also activated by contextually relevant body cues, suggesting its ability to integrate face and body signals.
Schmalzl et al. \shortcite{schmalzl_head_2012} found that selective activation patterns for consistent face and body information enable faster and more accurate responses through integrated perception.
Song et al. \shortcite{song_representation_2013} demonstrated that in the ventral visual pathway, signals related to faces and bodies are integrated in anterior regions, producing stronger neural responses compared to when these signals are processed separately in posterior regions.
Building on these psychological and neuroscience insights, we considered the possibility that \textcolor{rev}{misaligned} facial motion affects how observers perceive and understand a character.
\textcolor{rev}{To investigate this idea, we proposed the B2F framework and conducted a perceptual study, which provided supporting evidence.}

\paragraph*{Diverse Input-Based Motion Generation}

To create natural and expressive human motions, extensive research has focused on generating body and facial motions from diverse inputs such as video, images, speech, text, and music.
Video-based approaches aim to reconstruct 3D motions from the movements captured in input videos \cite{shimada_neural_2021,zhao_havatar_2023,danecek_emoca_2022},
while text-based methods generate motions by learning the abstract meanings of natural language \cite{temos,zhang_generating_2023,kim_flame_2023}.
Music-based studies produce dance movements driven by the rhythm and emotion of the music \cite{valle-perez_transflower_2021,alexanderson_listen_2023,zhuang_music2dance_2022},
and speech-based approaches use non-verbal cues like intonation, rhythm, and emotion to generate gestures or facial motions \cite{fan_faceformer_2022,aneja_facetalk_2024,danecek_emotional_2023}.
There has also been progress in simultaneously generating facial and full-body motions from various inputs \cite{zhao_media2face_2024,ao_gesturediffuclip_23,liu_emage_2024}.
\textcolor{ys}{While these studies focus on mapping from non-motion inputs (e.g., text, audio, video) to motion, our work takes a different direction: we treat body motion—typically regarded as an output—as the input, and generate the corresponding facial motion. Unlike speech-to-face animation methods which rely on explicit phoneme-to-lip correspondences, our approach leverages full-body motion to produce semantically aligned facial expressions, with additional control through a style reference. This defines a distinct problem setting with fundamentally different inputs and objectives.}

\paragraph*{Learning Temporal Semantic Correlations for Motion Generation}

\textcolor{rev}{
Research on learning temporal and semantic correlations for motion generation has primarily focused on mapping speech to facial expressions or gestures.
Recent models based on diffusion or transformers have effectively captured speech-motion alignment and enabled controllable, high-quality motion synthesis \cite{alexanderson_listen_2023,ao_gesturediffuclip_23,chhatre_emotional_2024,sun_diffposetalk_2024}.
Variational autoencoders (VAEs) have been used to produce diverse and detailed motions through probabilistic modeling \cite{ghorbani_zeroeggs_2023,pang_bodyformer_2023,danecek_emotional_2023}, while quantization-based methods have improved the handling of complex sequential dependencies \cite{ao_rhythmic_2022,ao_gesturediffuclip_23,xing_codetalker_2023,liu_emage_2024}.
Some recent studies have also explored simultaneously generating facial and full-body motions to enhance temporal and semantic coherence in character animation  \cite{chen_diffsheg_2024,yi_generating_2023,liu_emage_2024}
\textcolor{crr}{
, but they do not explicitly address body–face alignment, and their style control mechanisms are not tailored for this setting.
}
Building on these advances, we propose a new approach that models correlations between body and facial motions and enables dynamic style adaptation through user-specified references.
}

\paragraph*{Style Control in Motion Generation}

Controlling emotion- or identity-specific styles in generated facial or body motions is an important research topic, with various approaches proposed.
One approach is to use emotion or speaker identity labels as inputs. 
For instance, VOCA~\cite{cudeiro_capture_2019} and FaceFormer~\cite{fan_faceformer_2022} adjust speaking style based on speaker identity, while EMOTE~\cite{danecek_emotional_2023} and EmoTalk~\cite{peng_emotalk_2023} control emotional expression through emotion labels or intensity values.
Another approach is example-based control.
ZeroEGGS~\cite{ghorbani_zeroeggs_2023} generates gestures in the style of a given example, 
while GestureDiffuCLIP~\cite{ao_gesturediffuclip_23} uses prompts, such as text, motion, or video define stylistic attributes.
DiffPoseTalk~\cite{sun_diffposetalk_2024} generates gestures and facial motions based on styles derived from short reference videos, and
Motion Puzzle~\cite{jang_motion_2022} allows part-wise style control by combining different example motions.
Similarly, our method adopts an example-based approach, using  a facial style reference to control the style of the generated facial motion.

\textcolor{rev}{
One of the key challenges in style control is effectively disentangling style and content components in motion.
To address this, prior work has introduced various disentanglement techniques.
MeshTalk~\cite{richard_meshtalk_2021} introduced a cross-modality loss designed to separate upper and lower facial regions to independently reflect example motion and speech.
EmoTalk~\cite{peng_emotalk_2023} uses a cross-reconstruction loss by swapping content and emotion embeddings from speech, ensuring that generated facial motion reflects both factors.
AMUSE~\cite{chhatre_emotional_2024} extends this by disentangling emotion, content, and personal style through a similar loss formulation.
EMOTE~\cite{danecek_emotional_2023} proposed a disentanglement loss that compare swapped and non-swapped outputs using pretrained lip-reading and emotion feature extractors—removing the need for paired ground truth data.
Other approaches leverage consistency losses in learned embedding spaces to regularize content or style~\cite{danecek_emoca_2022, ao_gesturediffuclip_23}.
Inspired by these techniques, we train the B2F model end-to-end using (i) a consistency loss, (ii) a cross-consistency loss from swapping content and style inputs, and (iii) an alignment loss that connects body and facial content spaces.
\textcolor{ys}{This design enables not only effective disentanglement of content and style, but also maintains semantic coherence between generated facial motion and the driving body motion, while flexibly adapting to stylistic intent. In contrast to prior methods that often rely on curated paired data, pretrained modules, or multi-stage pipelines, our approach achieves these properties within an end-to-end unified generation framework, making it both practical and robust.}
}

\begin{figure*}[t]
    \centering
    \includegraphics[trim=0 0 0 0, clip, width=.85\textwidth]{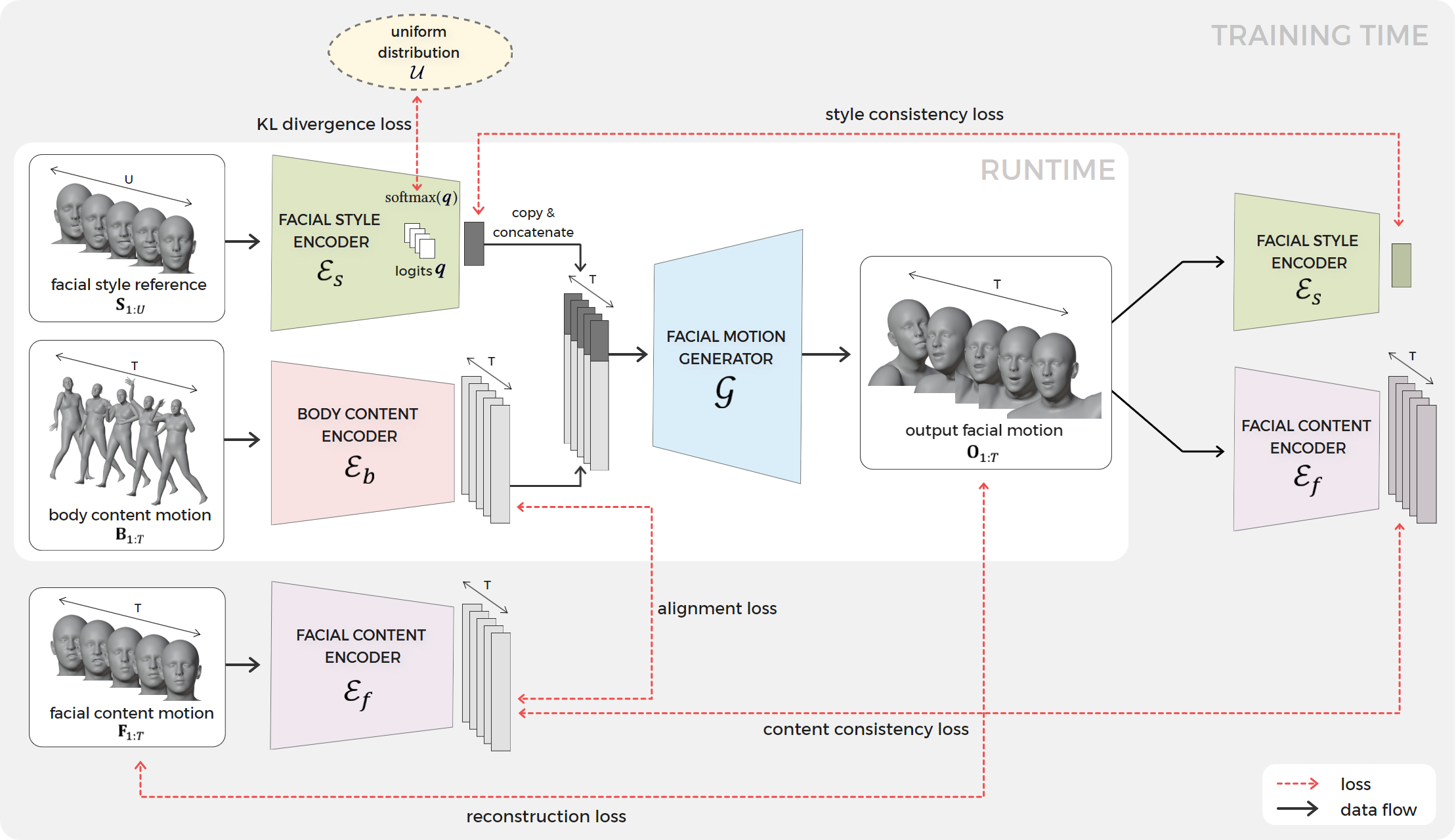}
    \caption{
    \textcolor{rev}{B2F architecture.
The B2F model consists of a body content encoder $\mathcal E_b$, facial content encoder $\mathcal E_f$, facial style encoder $\mathcal E_s$, and facial motion generator $\mathcal G$.
During training, it uses body and facial content motions along with a facial style reference to generate facial motion.
At inference, B2F generates facial motion using only the body content motion and style reference.
    }
    }
    \label{fig:architecture}
\end{figure*}

\section{B2F Architecture}

We propose a model that generates 3D facial motions aligned with an input body motion while reflecting a given facial style reference.
Our model consists a body content encoder $\mathcal E_b$, which maps the input body content motion $\mathbf B_{1:T}$ to a content embedding;
a facial content encoder $\mathcal E_f$, which maps the input facial content motion $\mathbf F_{1:T}$ to a content embedding; a facial style encoder $\mathcal E_s$, which maps the input facial style reference $\mathbf S_{1:U}$ to a style embedding; and a facial motion generator $\mathcal G$, which combines the content and style embeddings to produce the output facial motion $\mathbf O_{1:T}$, all trained end-to-end (Figure~\ref{fig:architecture}).

Each time step of the body motion $\mathbf B_{1:T}$ is defined as a set of positions, rotations, and linear velocities of key joints in the default gender-neutral SMPL-X model \cite{pavlakos2019expressive} at each frame.
Each time step of the facial motion $\mathbf F_{1:T}$, $\mathbf S_{1:U}$, and $\mathbf O_{1:T}$ is represented as a 53-dimensional FLAME~\cite{Li17} parameter vector, consisting of 50 expression parameters and 3 jaw pose parameters.
For details on the format of body content motion, please see Section 1 of the supplementary document.

\subsection{Facial Style Encoder}
\label{sec:facial-style-encoder}
{\color{rev}

The facial style encoder $\mathcal{E}_s$ takes the facial style reference $\mathbf{S}_{1:U}$ as input and outputs a fixed-dimensional style embedding $\mathbf{e}_s$.
To represent diverse facial styles while enabling differentiable training, we adopt a discrete latent variable structure using the Gumbel-Softmax trick~\cite{jang2017categorical}.
The input sequence is first processed by a Transformer-based encoder to produce logits for $D$ categorical distributions.
From each distribution, we sample a soft categorical vector via the Gumbel-Softmax trick, which approximates a discrete one-hot vector while allowing gradient flow.
These $D$ vectors of dimension $K$ are then flattened into a single style embedding $\mathbf{e}^\mathrm{s} \in \mathbb{R}^{D \cdot K}$.
This design balances distinctiveness and continuity in style representation.
While VAEs or VQ-VAEs can also model stylistic variation, this formulation was found to be more effective in producing consistent and expressive results in our setting.
See Section 2 of the supplementary document for further details.
}

\subsection{Body and Facial Content Encoders}

The body and facial content encoders, $\mathcal E_b$ and $\mathcal E_f$, process $\mathbf B_{1:T}$ and $\mathbf F_{1:T}$, respectively, to produce the body and facial content embeddings.
Both encoders are based on the same transformer-based architecture and are trained to extract shared content information.
During inference, $\mathcal E_b$ is used to generate content embeddings, while $\mathcal E_f$ is not utilized.
Please refer to Section 3 of the supplementary document for further details.

\subsection{Facial Motion Generator}

The facial motion generator $\mathcal G$ takes the style embedding from $\mathcal E_s$ and the content embedding from $\mathcal E_b$ as inputs to produce the output facial motion $\mathbf O_{1:T}$.
The style embedding is repeated for the sequence length $T$ and concatenated with the per-frame content embedding to form the input vectors. 
Based on a transformer decoder architecture, 
$\mathcal G$ is trained to generate facial motions that effectively capture the content and reflect the style from both embeddings.
Please refer to Section 4 of the supplementary document for further details.

\section{Training}

\subsection{Dataset and Batch Preparation}

To enable our B2F model to extract shared content information from body and facial motions, we utilized approximately 57 hours of data from the Motion-X dataset \cite{Lin23}, which captures facial and body motions simultaneously.
For more details about the dataset, please refer to Section 5 of the supplementary document.

The training data is structured as 180-frame body and facial motion segments, each with a 60-frame overlap.
During training, a batch is composed of $n$ pairs of body and facial motion segments randomly sampled from the training data.
For the selected $i$-th pair, a body content motion $\mathbf B_{1:T}^i$ is extracted from the body motion segment, while a facial content motion $\mathbf F_{1:T}^i$ and facial style reference $\mathbf S_{1:U}^i$ are extracted from the facial motion segment.
\textcolor{rev}{
The sequences $\mathbf{S}_{1:U}^i$, $\mathbf{F}_{1:T}^i$, and $\mathbf{B}_{1:T}^i$ are all sampled from the same starting frame within the motion segment.  
During training, the sequence lengths $U$ and $T$ are set to the same value to simplify the learning process, with $T$ randomly chosen between 60 and 90 frames. 
At inference time, however, the model can flexibly handle inputs where the facial style reference length $U$ and the body motion length $T$ are different.}
During training, two independent batches, batch A and batch B, are sampled in different random orders from the training data for loss computation. 
This means that the $\mathbf B_{1:T}$, $\mathbf F_{1:T}$, and $\mathbf S_{1:U}$ in batch A are derived from motion segment pairs that are distinct from those in batch B.

\subsection{Loss}
The model are trained using the following loss function:
\begin{equation}
\mathcal{L} = \lambda_1 \mathcal{L}_{\text{recon}} + \lambda_2 \mathcal{L}_{\text{align}} +  \lambda_3 \mathcal{L}_{\text{KL}}+ \lambda_4 \mathcal{L}_{\text{consi}} + \lambda_5 \mathcal{L}_{\text{cross}}.
\label{eq:total-loss}
\end{equation}

\begin{figure*}[t]
    \centering
    \includegraphics[trim=0 5 0 0, clip, width=.85\textwidth]{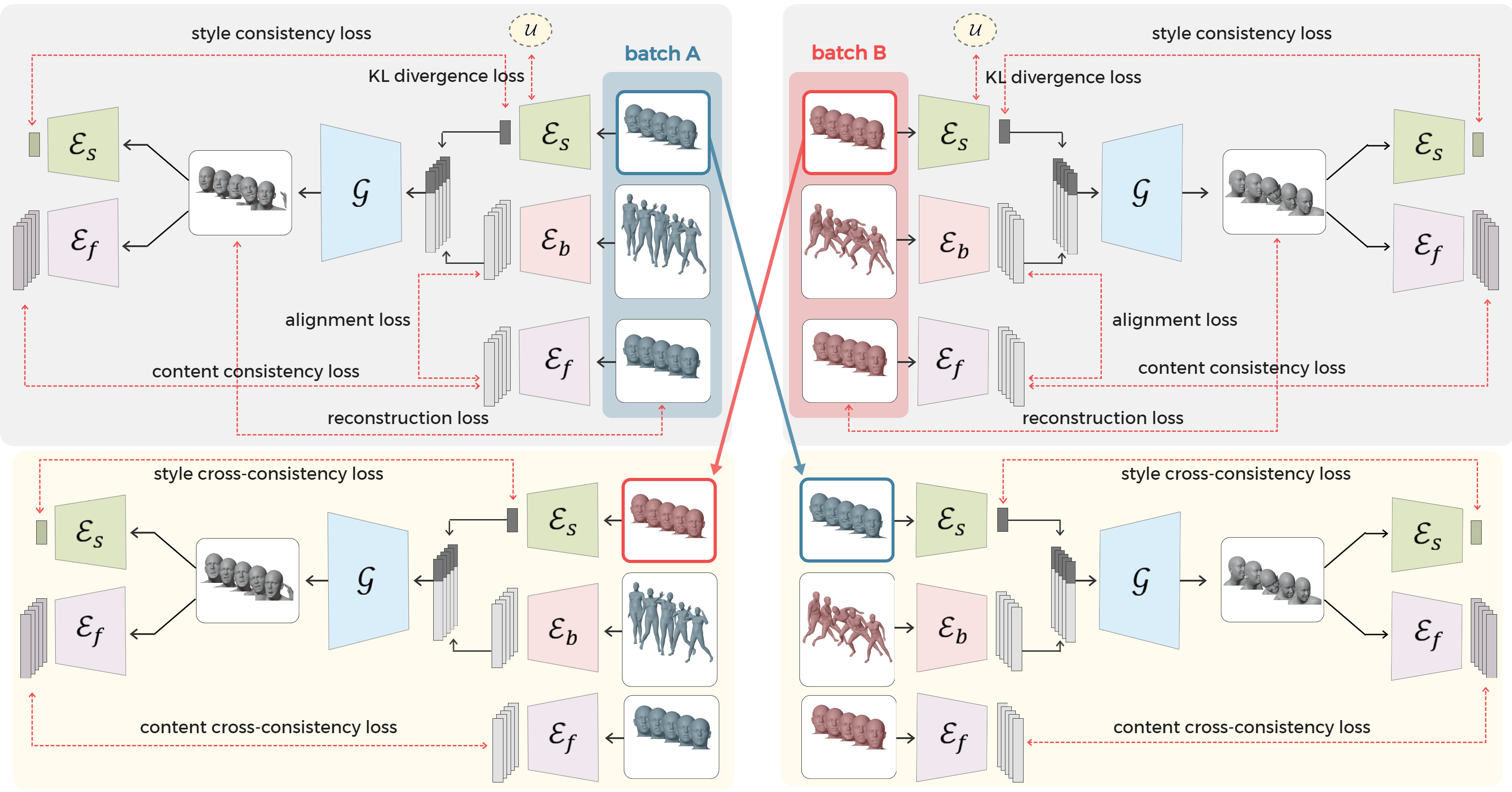}
    \caption{
    \textcolor{rev}{
    Loss computation.
    $\mathcal{L}_{\text{recon}}$, $\mathcal{L}_{\text{align}}$, $\mathcal{L}_{\text{KL}}$ and $\mathcal{L}_{\text{consi}}$ are calculated using the content and style input extracted from the same whole-body motion segment,
    while $\mathcal{L}_{\text{cross}}$ is calculated using the content inputs from one motion segment and the style input from a different segment.
    }
    }
    \label{fig:loss}
\end{figure*}

During training, the loss is first calculated and the network is updated using batch A, followed by a second loss calculation and update using batch B.
For simplicity, however, this section describes only the loss calculation based on batch A.
In this case, Equation~\ref{eq:total-loss} is computed using $\mathbf B_\textrm A$, $\mathbf F_\textrm A$, and $\mathbf S_\textrm A$ from batch A, along with $\mathbf S_\textrm B$ from batch B (Figure~\ref{fig:loss}).

\textbf{Reconstruction loss}
$\mathcal{L}_{\text{recon}}$ encourages the B2F model to reproduce the input facial content $\mathbf F_\textrm A$ when provided with the content and style inputs from batch A—namely, $\mathbf B_\textrm A$, $\mathbf F_\textrm A$, and $\mathbf S_\textrm A$, all extracted from the same whole-body motion segment:
\begin{equation}
\mathcal{L}_{\text{recon}} = \text{MSE}_\text{recon} \left( \mathbf F_\textrm A, \text{B2F}(\mathbf B_\textrm A, \mathbf F_\textrm A, \mathbf S_\textrm A) \right),
\end{equation}
where
$\text{B2F}(\cdot)$ represents the entire B2F process. %
For additional details, please refer to Section 6 of the supplementary document.

\textcolor{rev}{
\textbf{Alignment loss} $\mathcal{L}_{\text{align}}$ promotes alignment between the body and facial content embeddings:
\begin{equation}
\mathcal{L}_{\text{align}} = 
1 - \mathcal{E}_b(\mathbf{B}_\text{A}) \cdot \mathcal{E}_f(\mathbf{F}_\text{A}) .
\end{equation}
Since $\mathbf{B}_\textrm{A}$ and $\mathbf{F}_\textrm{A}$ are extracted from the same whole-body motion segment,  
$\mathcal{L}_{\text{align}}$ encourages $\mathcal{E}_b$ and $\mathcal{E}_f$ to extract consistent content representations from the body and facial motions at the same timestamps, measured via cosine similarity.
}

\textbf{KL divergence loss}  
$\mathcal{L}_{\text{KL}}$ regularizes each categorical distributions generated by the style encoder to approximate a uniform distribution, promoting balanced usage of all categories.
Given a style input $\mathbf{S}_\mathrm{A}$, the encoder outputs $D$ logit vectors, each of which is transformed into a probability distribution $q_d \in \mathbb{R}^K$ via the softmax function.
The loss is computed as the average KL divergence between each $q_d$ and a uniform distribution $\mathcal{U} \in \mathbb{R}^K$:
\begin{equation}
\mathcal{L}_{\text{KL}} = \frac{1}{D} \sum_{d=1}^{D} \text{KL}\left( q_d \,\|\, \mathcal{U} \right), \quad
q_d = \text{softmax}(\text{logits}_d(\mathbf{S}_\mathrm{A})).
\end{equation}
The term $q_d$ refers to the $d$-th categorical distribution from the style encoder given input $\mathbf{S}_\mathrm{A}$,
and $\text{logits}_d (\mathbf{S}_\mathrm{A})$ denotes the corresponding unnormalized logit vector.

\textbf{Consistency loss} $\mathcal{L}_{\text{consi}}$ ensures that the output facial motion faithfully reflects the input content and style. 
This is achieved by comparing the embeddings obtained by passing the output, generated from batch A's content and style inputs, through $\mathcal E_f$ and $\mathcal E_s$ with the embeddings obtained by passing the original inputs through the same encoders:
\begin{equation}
\mathcal{L}_{\text{consi}} = \mathcal{L}_{\text{consi}}^\text{content} + \mathcal{L}_{\text{consi}}^\text{style},
\end{equation}
\begin{equation}
\mathcal{L}_{\text{consi}}^\text{content} = 
\text{L2} \left( \mathcal E_f(\mathbf F_\text A), \mathcal E_f(\text{B2F}(\mathbf B_\textrm A, \mathbf F_\textrm A, \mathbf S_\textrm A) ) \right),
\end{equation}
\begin{equation}
\mathcal{L}_{\text{consi}}^\text{style} = 
\text{L2} \left( \mathcal E_s(\mathbf S_\text A), \mathcal E_s(\text{B2F}(\mathbf B_\textrm A, \mathbf F_\textrm A, \mathbf S_\textrm A) ) \right).
\end{equation}

\textbf{Cross-consistency loss} $\mathcal{L}_{\text{cross}}$ follows the same structure as $\mathcal{L}_{\text{consi}}$, with the key difference being that it incorporates the facial style from a different batch, "crossed" for the computation (Figure~\ref{fig:loss}).
Specifically, it compares the embeddings obtained by passing the output, generated from batch A's content inputs and batch B's style inputs, through $\mathcal E_f$ and $\mathcal E_s$ with the embeddings obtained by passing the original inputs through the same encoders:
\begin{equation}
\mathcal{L}_{\text{cross}} = \mathcal{L}_{\text{cross}}^\text{content} + \mathcal{L}_{\text{cross}}^\text{style},
\end{equation}
\begin{equation}
\mathcal{L}_{\text{cross}}^\text{content} = 
\text{L2} \left( \mathcal E_f(\mathbf F_\text A), \mathcal E_f(\text{B2F}(\mathbf B_\textrm A, \mathbf F_\textrm A, \mathbf S_\textrm B) ) \right),
\end{equation}
\begin{equation}
\mathcal{L}_{\text{cross}}^\text{style} = 
\text{L2} \left( \mathcal E_s(\mathbf S_\text B), \mathcal E_s(\text{B2F}(\mathbf B_\textrm A, \mathbf F_\textrm A, \mathbf S_\textrm B) ) \right).
\end{equation}

\textcolor{crr}{Using $\mathcal{L}_{\text{cross}}$ and $\mathcal{L}_{\text{consi}}$, the B2F model directly enforces that the content and style embeddings extracted from the generated facial motion match those from the input motions, even when the content or style inputs are cross-swapped. This loss design enables end-to-end training without pretrained components or datasets containing explicitly paired style–content combinations, while ensuring faithful style reflection and clear content–style disentanglement.}

\section{Extension for Various Animated Characters}
\label{sec:extension}

\begin{figure}
    \centering
    \includegraphics[trim=0 45 0 45, clip, width=1.\linewidth]{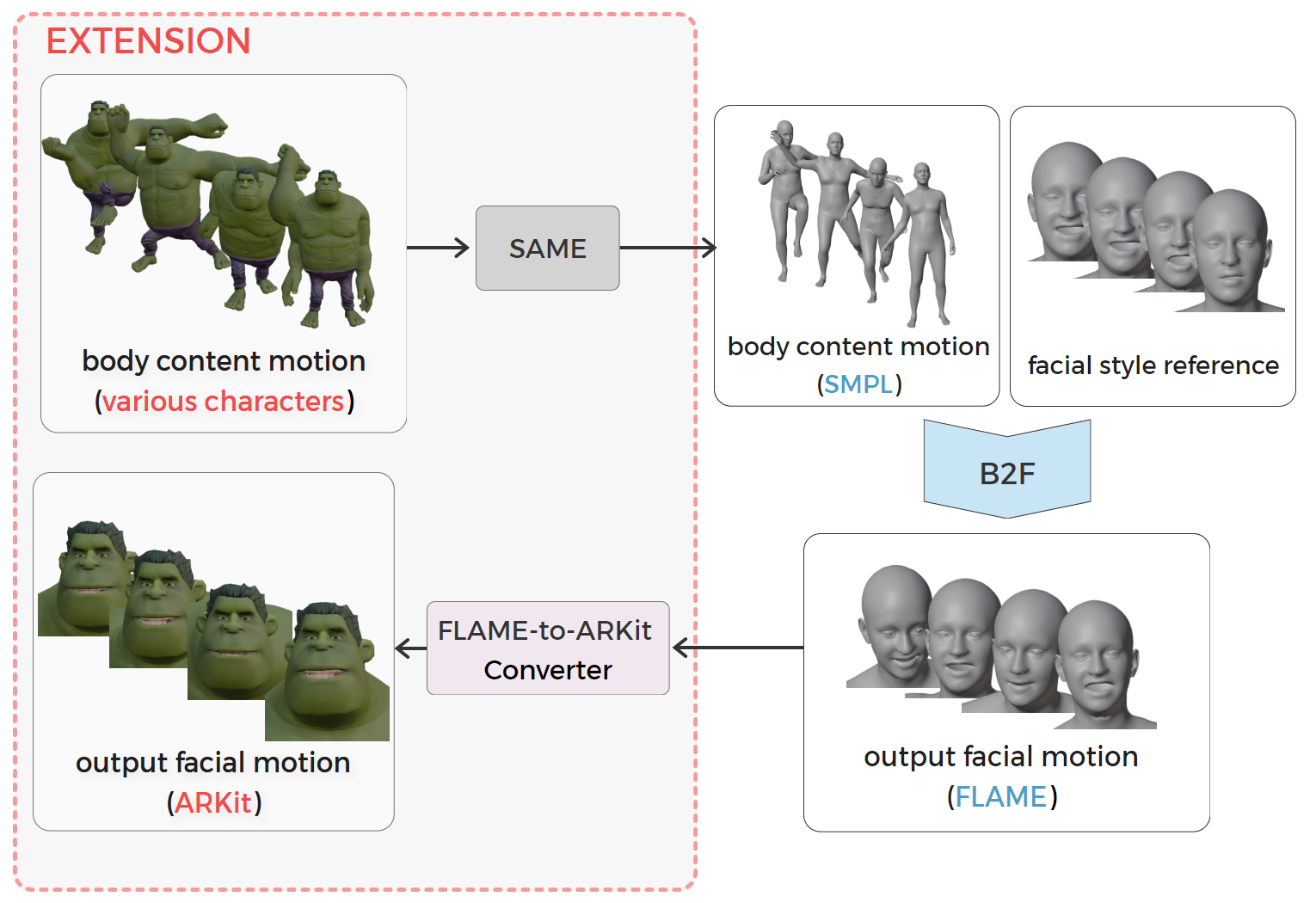}
    \caption{Extensions for various animated characters.
    }
    \label{fig:extension}
\end{figure}

\begin{figure*}
    \centering
    \includegraphics[trim=0 0 0 0, clip, width=.95\linewidth]{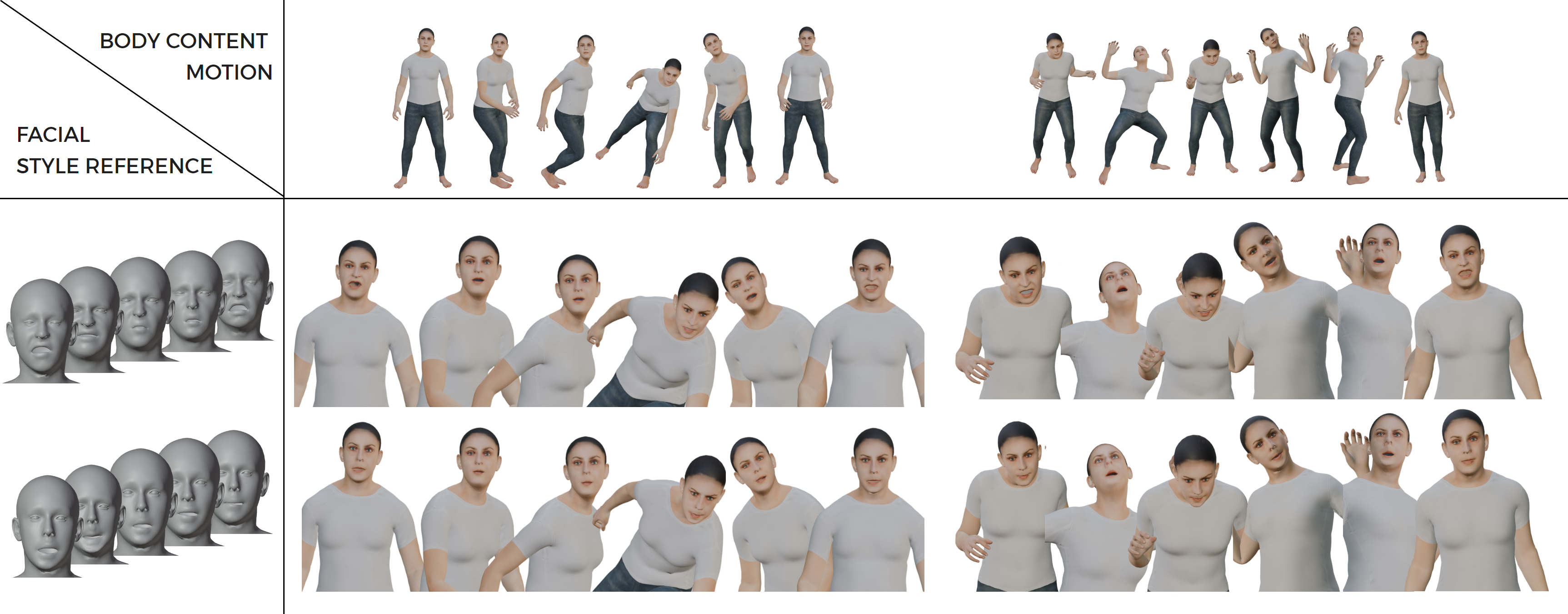}
    \caption{
    Results for various content and style inputs.
    }
    \label{fig:result-content-style}
\end{figure*}

{\color{rev}
The FLAME-format facial motion generated by B2F can be directly applied to SMPL-X characters but is less compatible with the diverse designs commonly used in animation and games.
To improve compatibility, we extend B2F with two modules: a motion retargeting module for the input and a FLAME-to-ARKit converter for the output (Figure~\ref{fig:extension}).

\paragraph*{Motion Retargeting for Input}
We use SAME~\cite{Lee23} to map body motion from various skeletons into a skeleton-agnostic embedding, which is then retargeted to SMPL-X’s default gender-neutral model. This enables B2F to accept diverse character motions as input.

\paragraph*{FLAME-to-ARKit Converter for Output}
To support more practical deployment, we introduce a FLAME-to-ARKit converter that maps FLAME parameters to iPhone ARKit blendshapes, a widely used format for facial animation in games and virtual avatars.
This module adopts a parameter-blended MoE architecture and is trained independently using ARKit facial motions from the BEAT dataset~\cite{liu_beat_2022} and their FLAME counterparts, converted via the transformation matrix from~\cite{liu_emage_2024}.
See Supplementary Section 7 for details.
}

\section{Qualitative Results}

\textcolor{rev}{
We qualitatively demonstrate the effectiveness of our method through diverse examples. The results are best observed in the accompanying video, and implementation details are provided in Supplementary Section~8.
}

\paragraph*{Various Content and Style Inputs}

\textcolor{rev}{
Our B2F model generates facial motions from unseen body content motions and facial style references not used during training.  
Figure~\ref{fig:result-content-style} shows results from various combinations of body content and facial style inputs.
}

\paragraph*{Various Animated Characters}

\begin{figure}
    \centering
    \includegraphics[trim=0 300 0 400, clip, width=0.8\linewidth]{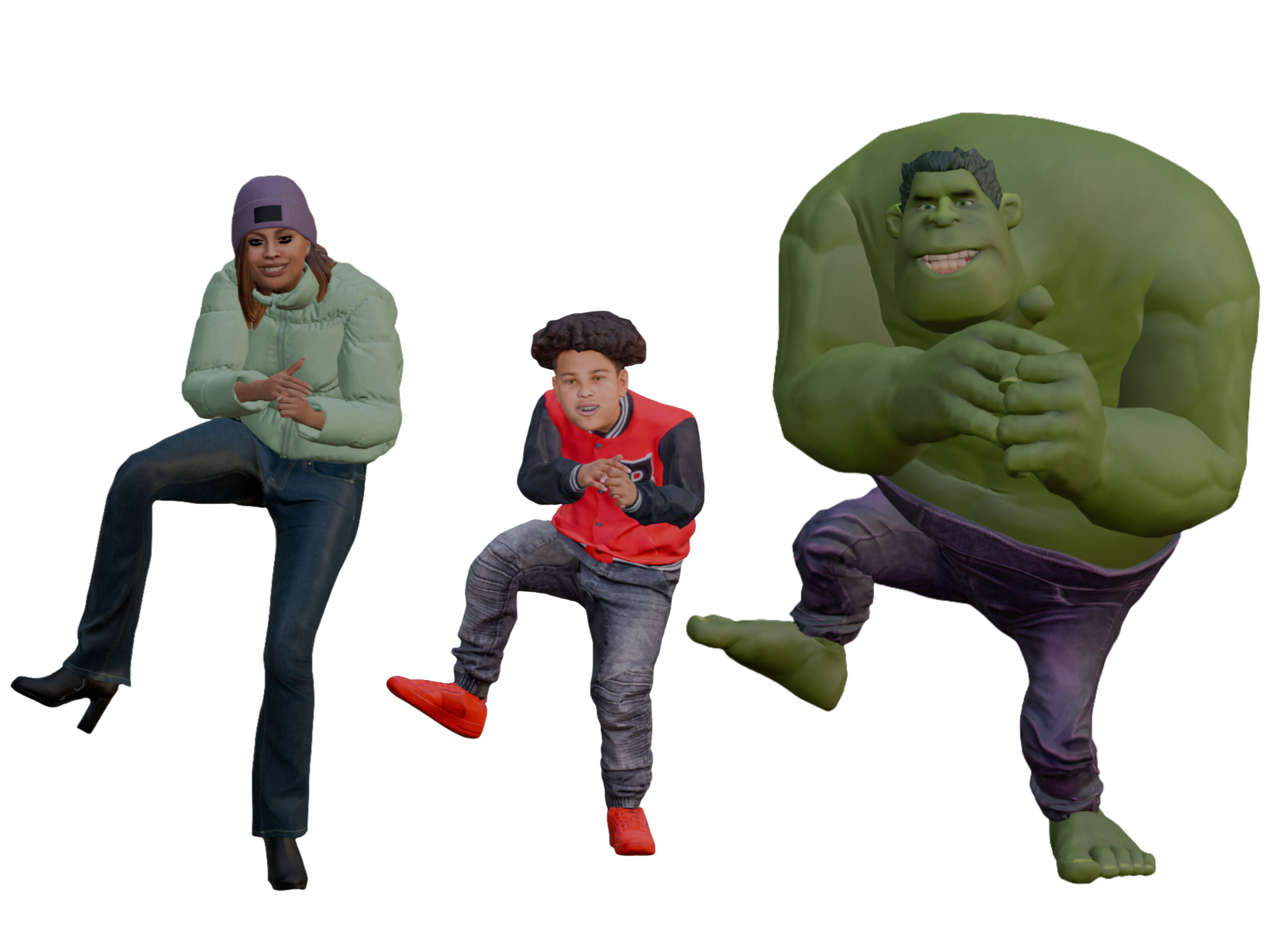}
    \caption{
    Results for various animated characters.
    }
    \label{fig:result-various-characters}
\end{figure}

\begin{figure}
    \centering
    \includegraphics[trim=0 0 0 0, clip, width=1.0\linewidth]{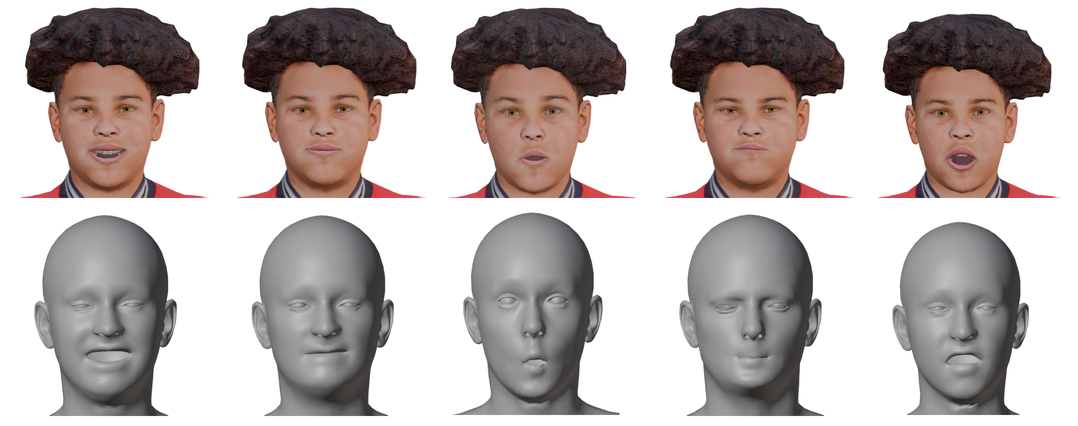}
    \caption{
    Examples of original FLAME facial motion (bottom) and converted ARKit facial motion (top). %
    }
    \label{fig:results-converter}
\end{figure}

\textcolor{rev}{We applied B2F to both stylized and realistic characters using paid and free assets from Reallusion’s ActorCore platform, which support 52 ARKit blendshapes.  
Figure~\ref{fig:result-various-characters} shows the results on various characters, and Figure~\ref{fig:results-converter} presents examples converted by our FLAME-to-ARKit converter.
}

\paragraph*{Style Interpolation and Transitions}

\begin{figure}
    \centering
    \includegraphics[trim=20 150 0 120, clip, width=1.\linewidth]{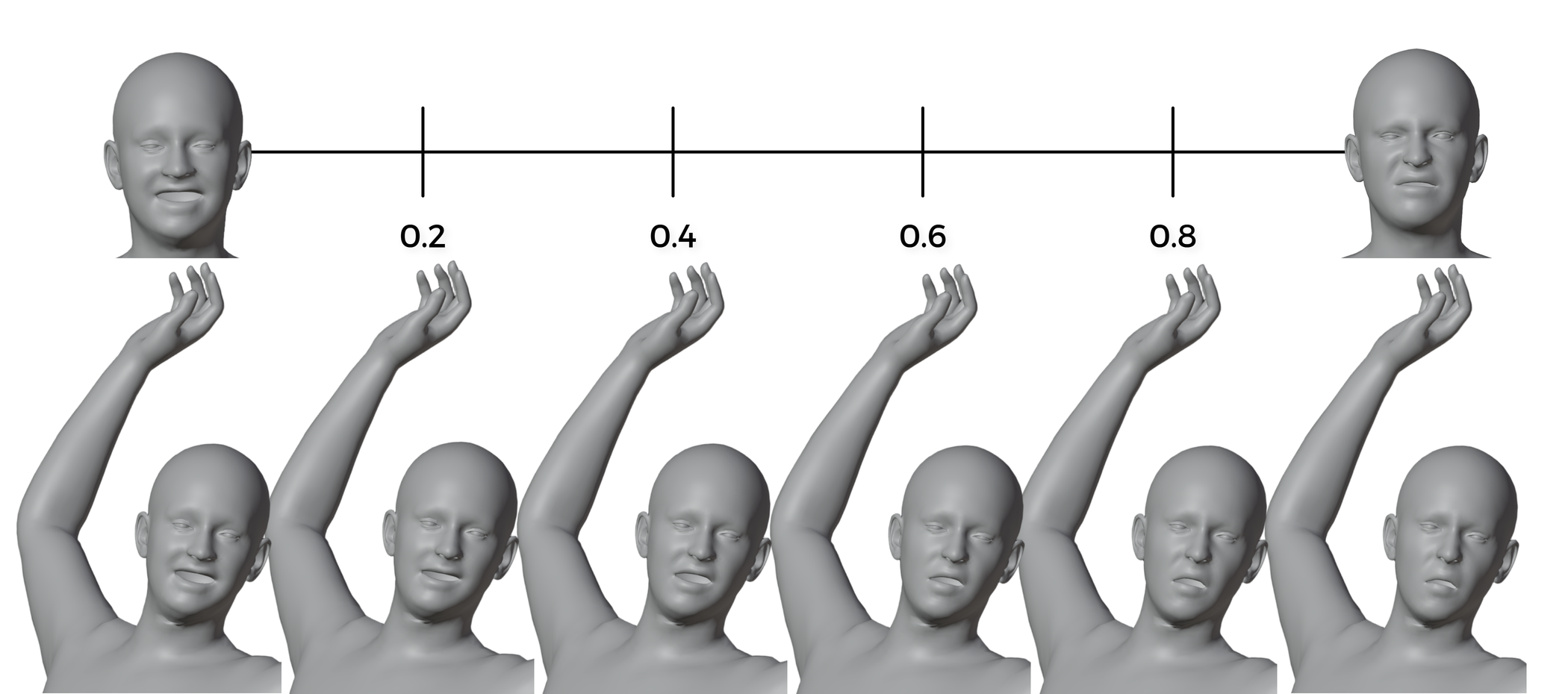}
    \caption{
    Results of style interpolation (ratios indicated). 
    }
    \label{fig:results-style-interp}
\end{figure}

\textcolor{rev}{The facial style encoder maps a style reference to an embedding, enabling style interpolation by blending embeddings from different references (Figure~\ref{fig:results-style-interp}).
By varying the interpolation ratio over time, our method also supports smooth style transitions during animation.}

\begin{figure*}
    \centering
    \includegraphics[trim=0 240 0 80, clip, width=.99\linewidth]{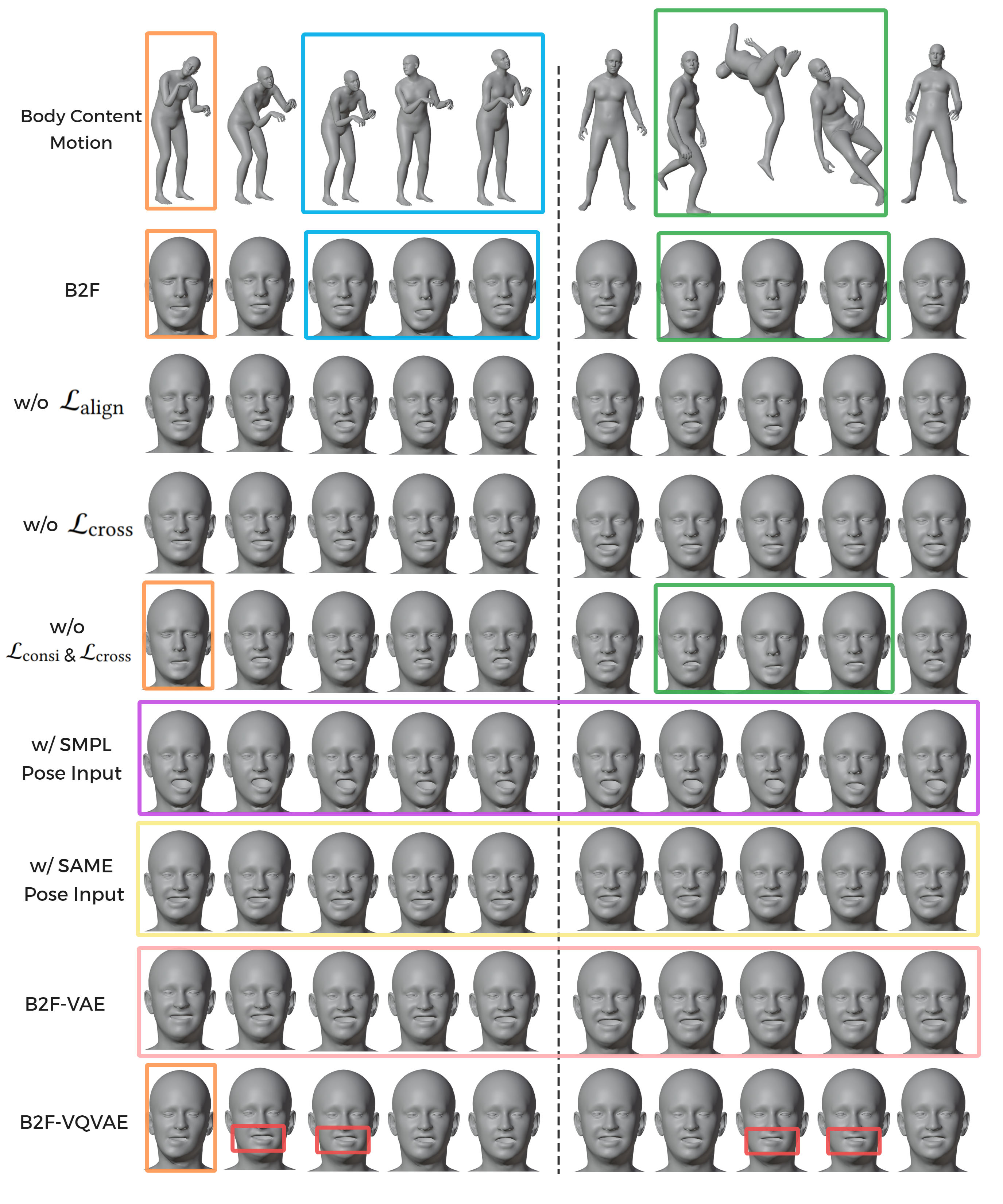}
    \caption{
    Ablation results across different body motions.
\textbf{Blue:} For rapid head-turning motions, only B2F shows a dynamic facial expression change.
\textbf{Green:} For backflip motions involving physical exertion, both B2F and w/o $\mathcal{L}_{\text{consi}} \& \mathcal{L}_{\text{cross}}$ reflect appropriate facial tension, but B2F shows the clearest expression shift.
\textbf{Orange:} For cat-like gestures, only B2F, w/o $\mathcal{L}_{\text{consi}} \& \mathcal{L}_{\text{cross}}$, and B2F-VQVAE produce fitting expressions, with B2F being the most expressive.
\textbf{Purple:} Facial expressions remain unchanged, with persistent artifacts such as asymmetrical lip shapes.
\textbf{Yellow:} No noticeable facial variation.
\textbf{Pink:} B2F-VAE exhibits minor expression changes, but they are overly subtle and lack expressiveness.
\textbf{Red:} B2F-VQVAE frequently generates distorted   mouth shapes throughout the sequence, regardless of the intended style.
    }
    \label{fig:results-ablation-content}
\end{figure*}

\begin{figure*}
    \centering
    \includegraphics[trim=0 240 0 650, clip, width=1.\linewidth]{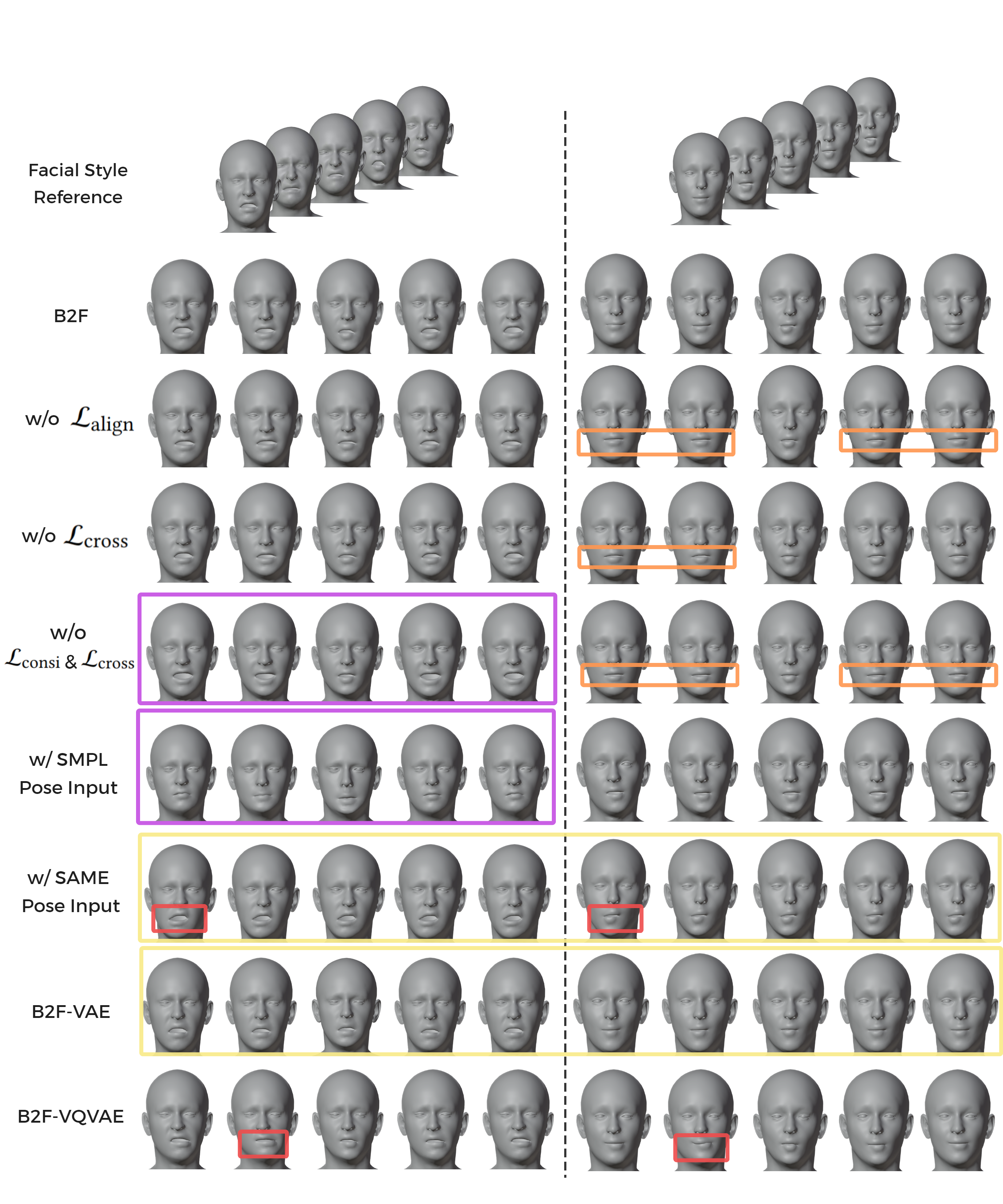}
    \caption{
    Ablation results across different facial style references.
\textbf{Purple:} Some models produce facial expressions that diverge from the intended style—e.g., w/o $\mathcal{L}_{\text{consi}} \& \mathcal{L}_{\text{cross}}$  yields positive expressions despite a negative style, and the SMPL Pose Input model shows tightly closed lips regardless of the input.
\textbf{Yellow:} Style is somewhat reflected, but facial expression changes are minimal.
\textbf{Red:} Distorted lips persist throughout the sequence; B2F-VQVAE shows particularly severe artifacts.
\textbf{Orange:} Despite the style input conveying tense lips, the generated expressions fail to retain lip tension.
    }
    \label{fig:results-ablation-style}
\end{figure*}

{\color{rev}
\section{Ablation Study}

To analyze the contributions of individual components in our B2F model, we conduct an ablation study with both quantitative and qualitative comparisons.
\textcolor{ys}{As no prior method directly addresses body-to-face motion generation with style control, ablation offers the most reliable means to isolate and assess the effectiveness of each design choice.}

\paragraph*{Compared Models}
In addition to the full \textbf{B2F (ours)} model, we evaluate the following ablation and alternative models:
\begin{itemize}[leftmargin=1.5em]
  \item \textbf{w/o $\mathcal{L}_{\text{align}}$}: model without the alignment loss.
  \item \textbf{w/o $\mathcal{L}_{\text{cross}}$}: model without the cross consistency loss.
  \item \textbf{w/o $\mathcal{L}_{\text{consi}} \& \mathcal{L}_{\text{cross}}$}: model without any consistency losses.
  \item \textbf{w/ SMPL Pose Input}: model that directly uses SMPL pose parameters as the body content encoder input.
  \item \textbf{w/ SAME Pose Input}: model that replaces the body content encoder with a pretrained SAME encoder.
  \item \textbf{B2F-VAE}: model using a Gaussian VAE instead of Gumbel-Softmax VAE.
  \item \textbf{B2F-VQVAE}: model using VQ-VAE for discrete latent representation instead of a Gumbel-Softmax-based VAE.
\end{itemize}

To ensure consistent evaluation, we used the mean of the Gaussian latent for B2F-VAE and a near-deterministic setting ($\tau = 10^{-6}$) for Gumbel-Softmax in B2F (ours).

\subsection{Quantitative Comparison}
We randomly sampled 300 clips from the Motion-X dataset for evaluation, totaling 78,185 frames (approx. 43 minutes). For each clip, facial motion was generated based on the entire body content and facial style input.
Evaluation was performed in the FLAME blendshape weight space, matching the model’s output representation.
We used two measures: (1) the average $\ell_2$ error between predicted and ground-truth blendshape weights to assess reconstruction accuracy, and (2) the average absolute difference in standard deviation across each blendshape dimension over time to assess temporal variation, inspired by the Facial Deformation Distance (FDD) metric~\cite{xing_codetalker_2023}.

\begin{table}[h]
\centering
\caption{Frame-wise $\ell_2$ error and standard deviation difference between predicted and ground-truth blendshape weights.}
\begin{tabular}{lcc}
\toprule
\textbf{Model} & $\ell_2$ Error $\downarrow$ & Std. Dev. Difference $\downarrow$ \\
\midrule
\textbf{B2F (ours)}            & 0.556 & 0.309 \\
\textbf{w/o $\mathcal{L}_{\text{align}}$}      & 0.593 & 0.331 \\
\textbf{w/o $\mathcal{L}_{\text{cross}}$}      & 0.591 & 0.326 \\
\textbf{w/o $\mathcal{L}_{\text{consi}} \& \mathcal{L}_{\text{cross}}$} & 0.565 & 0.297 \\
\textbf{w/ SMPL Pose Input}     & 0.618 & 0.381 \\
\textbf{w/ SAME Pose Input}     & 0.673 & 0.564 \\
\textbf{B2F-VAE}                & 0.567 & 0.321 \\
\textbf{B2F-VQVAE}              & 0.570 & 0.292 \\
\bottomrule
\end{tabular}
\label{tab:quantitative}
\end{table}

Table~\ref{tab:quantitative} summarizes the results. \textbf{B2F (ours)} achieves the lowest reconstruction error and one of the lowest scores in temporal variation error, indicating high accuracy and reliable expressiveness consistent with the ground-truth dynamics.
\textbf{w/ SMPL Pose Input} and \textbf{w/ SAME Pose Input} models perform the worst, showing the highest errors.
\textbf{w/o $\mathcal{L}_{\text{consi}} \& \mathcal{L}_{\text{cross}}$} shows low temporal deviation error and performs well in the current metric setting where style and content inputs are identical. However, it lacks the ability to capture fine-grained style details, leading to visually less expressive results.
\textbf{B2F-VAE} appears to suffer from blurry style boundaries and weaker expressiveness, possibly due to the difficulty of learning distinct styles in a fully continuous latent space. This may have contributed to its higher error in both accuracy and variation compared to \textbf{B2F (ours)}.
\textbf{B2F-VQVAE}, while numerically competitive, often displays unnatural expressions such as asymmetric lips due to the rigid discretization of style into a finite codebook.

\subsection{Qualitative Comparison}

We also qualitatively compared the models in terms of their ability to reflect both body content motion and facial style reference.

Figure~\ref{fig:results-ablation-content} shows facial expressions generated for different body motions. \textbf{B2F} produces clear, expressive changes such as tightly closed lips during high physical exertion and dynamic reactions to rapid head turns. In contrast, ablation models produce more muted expressions or display persistent artifacts, such as distorted mouths. The SMPL and SAME variants show little expression diversity.

Figure~\ref{fig:results-ablation-style} compares expression styles under different facial style references. \textbf{B2F} captures fine-grained style cues (e.g., subtle lip tension) across sequences. Alternative models, especially those without consistency loss or using discrete codebooks, either reflect the style incorrectly or exhibit strong visual artifacts like asymmetrical lips.
SMPL tends not to reflect stylistic variation effectively, and SAME shows limited expression changes with some distorted lip artifacts.

}

{\color{rev}

\section{Perceptual Study}

To evaluate how our model’s facial motion influences perception, we conducted two user studies on Amazon Mechanical Turk: one comparing it with a static face and another with a misaligned face.

\paragraph*{Experiment 1: B2F vs. Static Face}
Participants were shown side-by-side characters in a video, each performing the same body motion. One side included facial motion generated by our model (B2F), while the other displayed a static neutral face.
The two conditions were presented in random order, and participants were not informed about the specific difference between them.  
Participants were asked to evaluate each pair based on three aspects:  
1) \textit{"Which character's animation looks more natural \& realistic?"}  
2) \textit{"Which character's animation is more engaging \& immersive?"}  
3) \textit{"Which character's emotions \& intentions are expressed more clearly?"}  
Responses were recorded on a five-point Likert scale ranging from "Strongly A" to "Strongly B", with a neutral "No difference" option.

Each participant completed 10 trials, including 2 catch trials where they compared a step-motion character—with poses changing once per second and a static facial expression—to a fully animated sequence generated by B2F.
Participants who selected the step-motion condition as more realistic in catch trials were excluded.  
Responses from the 30 participants who passed the validation criteria were used for analysis.

\begin{figure}
    \centering
    \includegraphics[trim=25 70 25 70, clip, width=1.\linewidth]{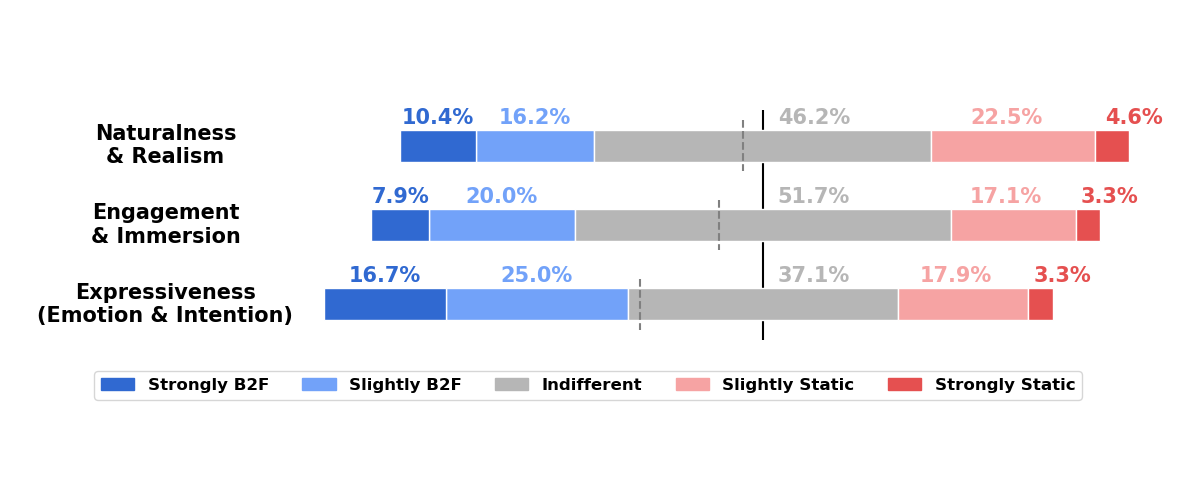}
    \caption{
    Results of Experiment 1: B2F vs. Static Face.
    }
    \label{fig:perceptual-static}
\end{figure}

As shown in Figure~\ref{fig:perceptual-static}, B2F was notably preferred over the static face in expressiveness, while preferences for naturalness and engagement were more balanced.
In terms of expressiveness (emotion \& intention), B2F received the strongest preference, with 16.7\% of responses indicating “Strongly B2F” and an additional 25.0\% choosing “Slightly B2F.”
For engagement \& immersion, the combined preference for B2F reached 27.9\%, although over half of participants (51.7\%) reported no clear difference.
For naturalness \& realism, responses were more mixed, with 26.6\% favoring B2F and 27.1\% favoring the static face. This suggests that while B2F enhances emotional and communicative aspects of animation, the perceived naturalness may not differ significantly from a static expression in all cases.

\paragraph*{Experiment 2: B2F vs. Misaligned Face}
In the second experiment, the setup was identical to Experiment 1 in terms of question phrasing, interface, and the number of participants (30).
However, the comparison conditions differed: one side showed a character with facial motion generated by B2F for the given body motion (as in Experiment 1), while the other side used the same body motion combined with facial motion generated by B2F from a different, mismatched body motion.
Specifically, for each of the 8 main questions, facial motions originally paired with other body motions were reassigned—for example, body motion 1 was paired with facial motion generated from body motion 2, body motion 2 with facial motion from body motion 5, and so on—thereby intentionally misaligning the body content and facial content.
A separate group of 30 participants was recruited for this experiment.

\begin{figure}
    \centering
    \includegraphics[trim=25 70 25 70, clip, width=1.\linewidth]{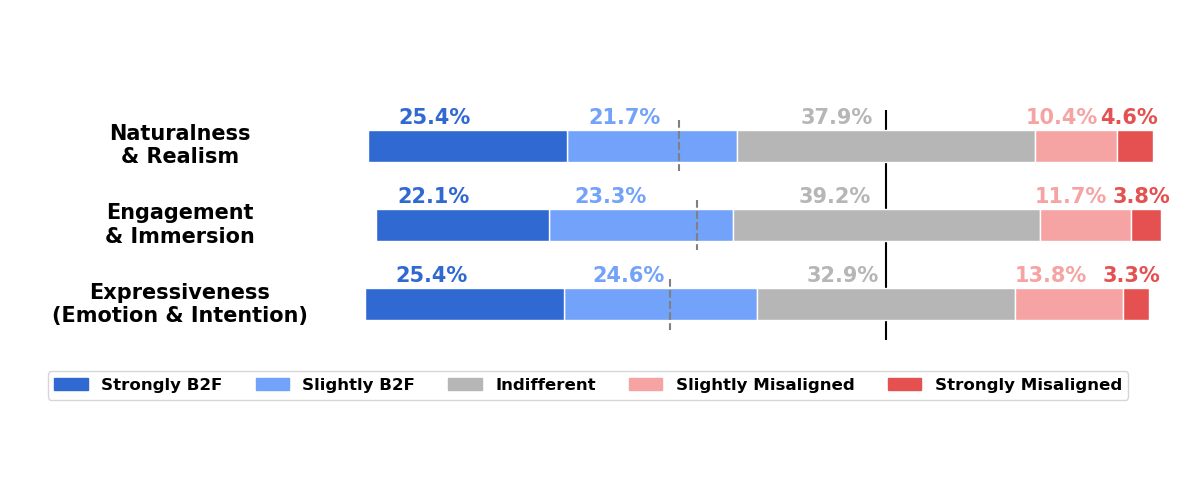}
    \caption{
    Results of Experiment 2: B2F vs. Misaligned Face.
    }
    \label{fig:perceptual-misaligned}
\end{figure}

As shown in Figure~\ref{fig:perceptual-misaligned}, the aligned B2F animation was clearly preferred in all three evaluation categories.
In expressiveness (emotion \& intention), a strong preference for B2F was observed, with 25.4\% of responses indicating “Strongly B2F” and 24.6\% selecting “Slightly B2F,” while only 17.1\% preferred the misaligned version.
Similar trends were observed for engagement \& immersion (22.1\% strongly B2F, 23.3\% slightly B2F) and naturalness \& realism (25.4\% strongly B2F, 21.7\% slightly B2F).
These results indicate that mismatched facial motions noticeably degrade the perceived quality of character animation, highlighting the importance of synchrony between body and facial cues.

These results suggest that mismatched facial motion may negatively impact perception even more than having no facial motion at all. While a static face lacks expressiveness, it does not appear to conflict with body movements. In contrast, facial expressions that are temporally or semantically inconsistent with the accompanying body motion could violate users’ expectations and potentially lead to perceptual dissonance.

}

\section{Real-time Performance Evaluation}

\textcolor{ys}{To assess the real-time feasibility of our system, we simulated a scenario where full-body motion was pre-recorded and played at 30 FPS in Blender, while our model generated facial motion in real time. At each frame, the system used the previous 50 frames of body motion as input to produce a facial motion sequence, applying the last frame as the current facial expression. This setup matches the 30 FPS training configuration of our B2F model, ensuring consistent temporal resolution. On an NVIDIA RTX 4090 GPU, the system achieved an average processing time of 10.3–13.3 ms per frame (equivalent to 75–96 FPS), while using about 52 MB of allocated GPU memory and up to 130 MB reserved. These results indicate that the system can stably generate facial animations alongside complex 3D graphics tasks, suggesting that full end-to-end real-time generation is theoretically feasible.}

\section{Discussion}

We introduced B2F, a method for generating facial motions aligned with body motions.
By disentangling content and style information, B2F creates context-appropriate facial expressions while reflecting a specified style.
The proposed FLAME-to-ARKit converter enables the application of B2F to a wide range of characters using ARKit blendshapes.

\textcolor{rev}{
Our perceptual studies showed that B2F was generally preferred over both static and misaligned alternatives, especially for expressiveness and engagement, with similar or slightly better results in realism. These findings highlight the importance of synchrony between facial and bodily cues and suggest that mismatched facial motion may degrade perceptual quality even more than the absence of facial motion.
This underscores B2F’s potential to mitigate perceptual dissonance caused by incongruent multimodal cues.
}

\textcolor{ys}{Beyond perceptual quality, B2F demonstrated notable generalization and flexibility. During training, body and style inputs of moderate length were used, yet the model produced stable facial motion for test sequences exceeding 5,000 frames—far outside the training range. This suggests that B2F can robustly handle long and variable body motion sequences without degradation. In addition, the Gumbel-Softmax–based latent structure of the style encoder allows for diverse and consistent style representations with continuous gradient flow. This design enables the model to capture intended styles even from exaggerated or unseen references, while maintaining plausible expression ranges and semantic coherence with the body motion. Together, these properties make B2F adaptable to a broad spectrum of motion and stylistic conditions.}

This research was motivated by the observation that using skinned mesh characters with static faces in studies might influence the perception of body motion due to the lack of facial expression dynamics.
While our study shows that dynamic facial animation may not always improve perceived realism compared to static faces, it can enhance expressiveness and engagement. By sharing our code upon acceptance, we aim to support researchers in augmenting their full-body motion presentations with dynamic and style-controllable facial animations, potentially enriching character-driven studies across various applications.

One limitation of our approach is the inability to generate eye blinking due to the lack of eye blinking motion in the dataset.
Addressing this limitation and incorporating realistic blinking patterns into our framework is an area of focus for future work.
Although the study involved a total of 60 AMT participants, future work could include more diverse participant demographics and repeated measures to assess intra-subject consistency.
\textcolor{rev}{
Furthermore, we plan to extend our framework to accept natural language descriptions of facial style, making it easier and more intuitive for users to control the style of generated facial animations.
}

\section*{Acknowledgments}
This work was supported by the National Research Foundation
of Korea (NRF) grant (RS-2023-00222776); and by Culture, Sports and
Tourism R\&D Program through the Korea Creative Content Agency
grant funded by the Ministry of Culture, Sports and Tourism in 2024
(RS-2024-00399136).

\bibliographystyle{eg-alpha-doi}
\bibliography{citation}

\clearpage
\appendix
\begin{strip}
\centering
{\fontsize{18pt}{21pt}\bfseries Supplementary Document for B2F: End-to-End Body-to-Face Motion Generation with Style Reference\par}
\vspace{20pt}
\end{strip}

\section{Body Content Motion Format}

Each time step $\mathbf b_t$ of the body content motion $\mathbf B_{1:T}$ is defined as:
$\mathbf b_t = (\mathbf p_t, \mathbf r_t, \mathbf v_t)$, where $\mathbf p_t$, $\mathbf r_t$, and $\mathbf v_t$ represent the positions, rotations, and linear velocities of key joint in the default gender-neutral SMPL-X model \cite{pavlakos2019expressive}.
These values are expressed in the character frame, with the origin at the root (pelvis) position projected onto the ground. The axes of the character frame are defined by the vector obtained by projecting the root's forward direction onto the ground, the global up vector, and their cross-product.

The key joints used are as follows: `pelvis', `left\_ankle', `right\_ankle', `left\_foot', `right\_foot', `head', `left\_elbow', `right\_elbow', `left\_wrist', and `right\_wrist'.

\section{Facial Style Encoder Architecture}

\begin{figure}
    \centering
    \includegraphics[width=0.7\linewidth]{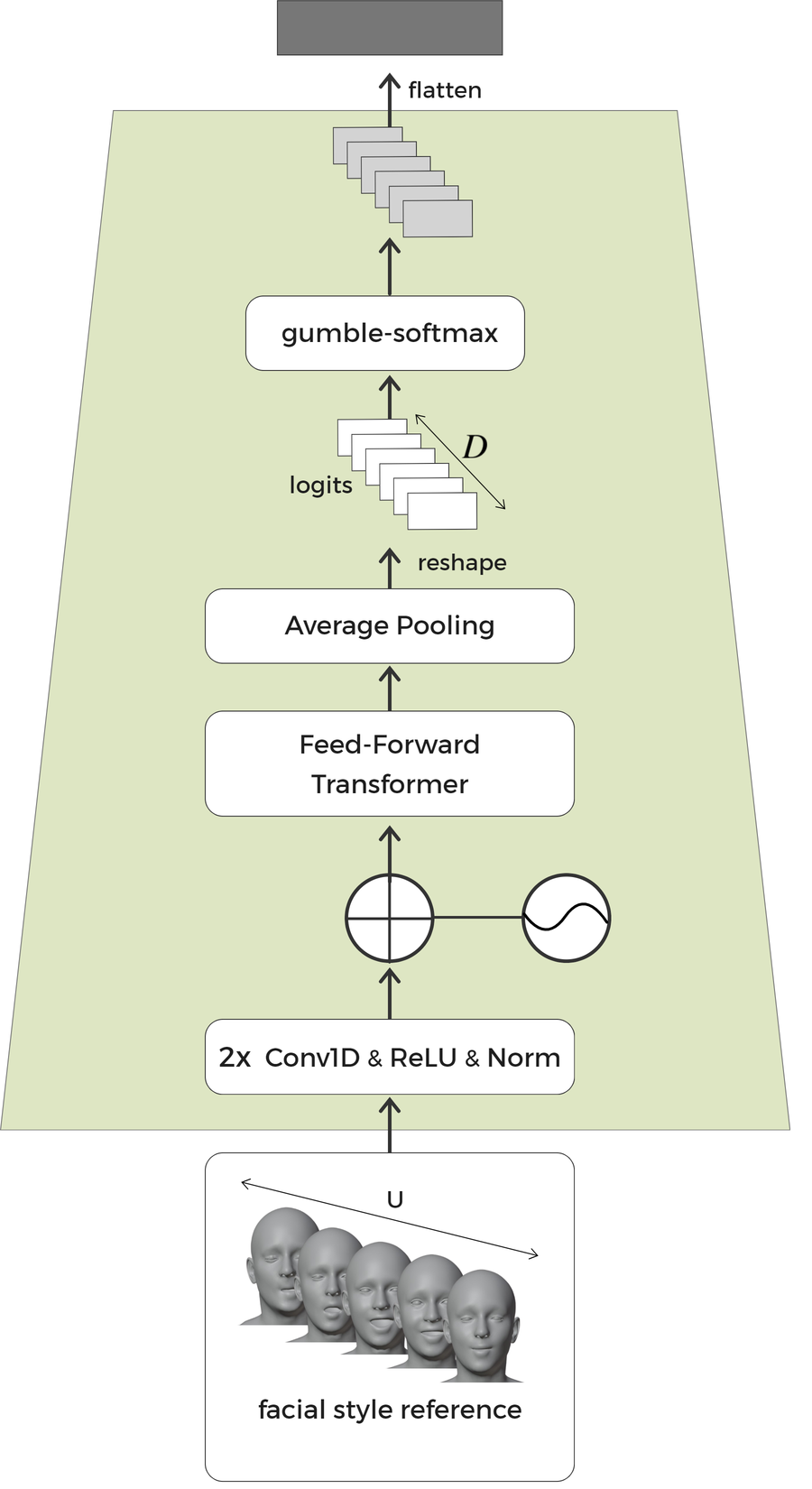}
    \caption{\color{rev}Facial Style Encoder architecture.}
    \label{fig:style-encoder}
\end{figure}

{\color{rev}{
The facial style encoder adopts a transformer-based architecture to encode variable-length style reference sequences into discrete latent vectors. The output sequence is mean-pooled across time and converted into categorical logits, which are transformed into a continuous approximation of one-hot vectors via the Gumbel-Softmax trick \cite{jang2017categorical}.

The facial style reference $\mathbf{S}_{1:U}$ is first processed through two consecutive 1D convolutional layers with kernel size 3 and padding 1. Each layer is followed by ReLU activation and layer normalization. A sinusoidal positional encoding is then added to retain the temporal order of the input sequence.
The encoded sequence is then passed through a Feed-Forward Transformer block \cite{Zaidi2021}, consisting of a multi-head self-attention layer and two position-wise 1D convolutional layers. Each component is followed by residual connections and layer normalization. This produces a sequence of hidden vectors with shape $[B, U, H]$, where $B$ is the batch size and $H = D \cdot K$.
Average pooling is applied over the temporal dimension to obtain a fixed-size representation of shape $[B, D \cdot K]$, which is reshaped into categorical logits of shape $[B, D, K]$. In our implementation, we set the number of categorical variables to $D = 12$ and the number of categories per variable to $K = 16$.

To enable differentiable sampling, Gumbel noise is added to each categorical logit, and a softmax operation with temperature $\tau$ is applied. This produces a soft approximation of a one-hot vector for each categorical variable. During training, $\tau$ is fixed to 0.7 to balance between discreteness and gradient stability. Optionally, hard sampling can be used during inference by selecting the maximum index. The final style embedding $\mathbf{e}^\mathrm{s} \in \mathbb{R}^{B \times D \cdot K}$ is obtained by flattening the $[B, D, K]$ tensor.
}}

\section{Body and Facial Content Encoder Architecture}

The Body Content Encoder and Facial Content Encoder share an identical architecture based on a Transformer encoder. Each processes the input sequence using 4 attention heads and 8 Transformer layers to effectively learn and extract content information. A subsequent linear layer maps the output to the final embedding dimension, followed by normalization to ensure a unified scale for the content embeddings.

Both encoders generate 512-dimensional embeddings for each frame of the input sequence. The Body Content Encoder has an input size of 144 per frame, while the Facial Content Encoder has an input size of 53 per frame.

\begin{figure}
    \centering
    \includegraphics[width=0.7\linewidth]{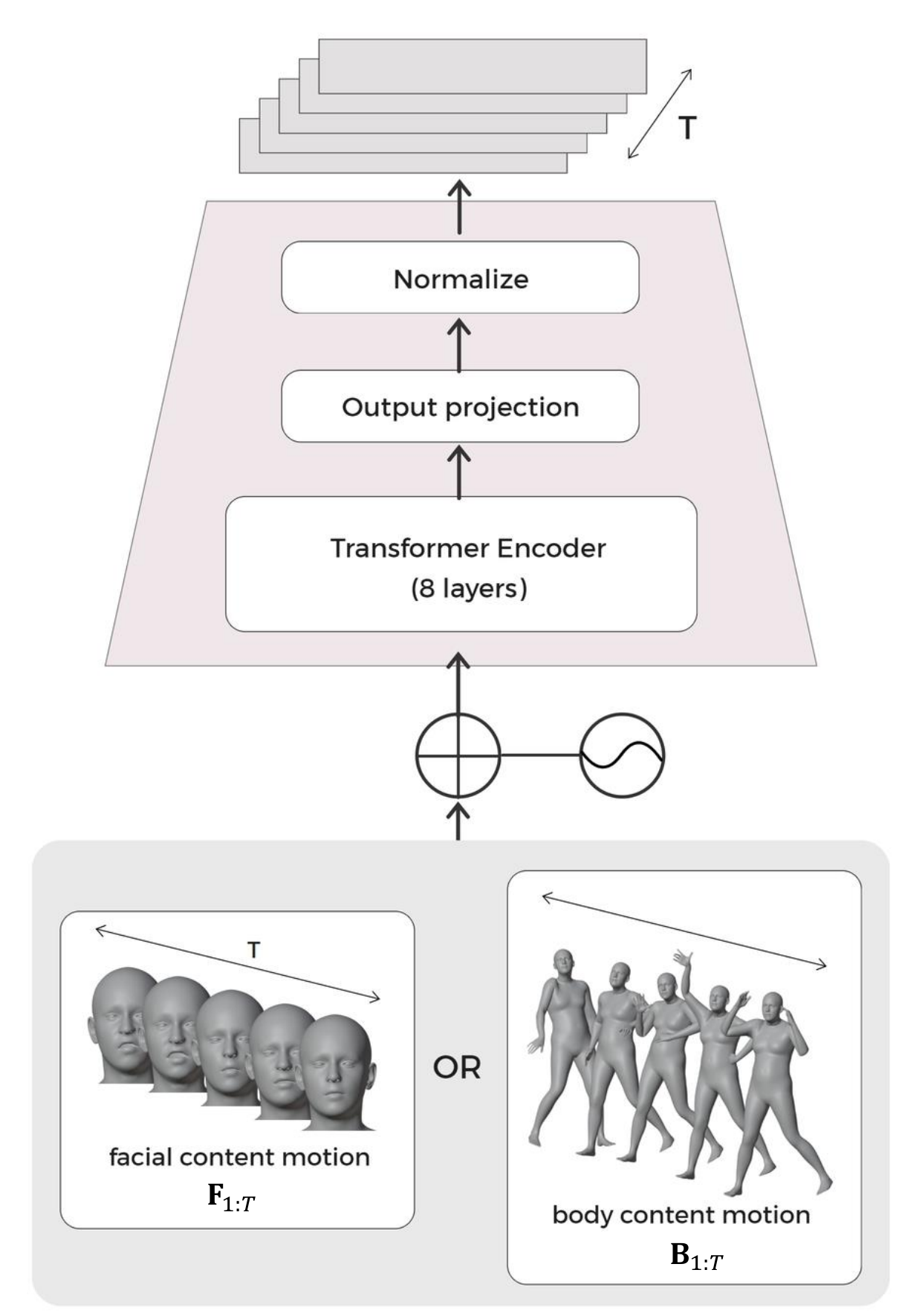}
    \caption{Body and Facial Content Encoder architecture.}
    \label{fig:content-encoder}
\end{figure}

\section{Facial Motion Generator Architecture}

{\color{rev}{
The Facial Motion Generator is built upon a Transformer decoder architecture, designed to generate temporally coherent and stylistically expressive facial animations. The body content encoder extracts a sequence of content embeddings from the input body motion, which serves as the memory of the Transformer decoder.
To inject style information, the style embedding produced by the facial style encoder is first projected and then broadcast along the time axis to match the sequence length. This time-aligned style embedding is concatenated with the body content embedding at each frame and used as the input to the Transformer decoder.

In this architecture, the decoder memory contains only the content embedding, allowing the model to attend to body motion features when generating facial expressions. Style information is not used as memory but instead conditions the decoding process through the concatenated input. The output of the decoder is then passed through a linear mapping layer to produce the final frame-wise 53-dimensional facial blendshape sequence in FLAME format.
}}

\begin{figure}
    \centering
    \includegraphics[width=0.7\linewidth]{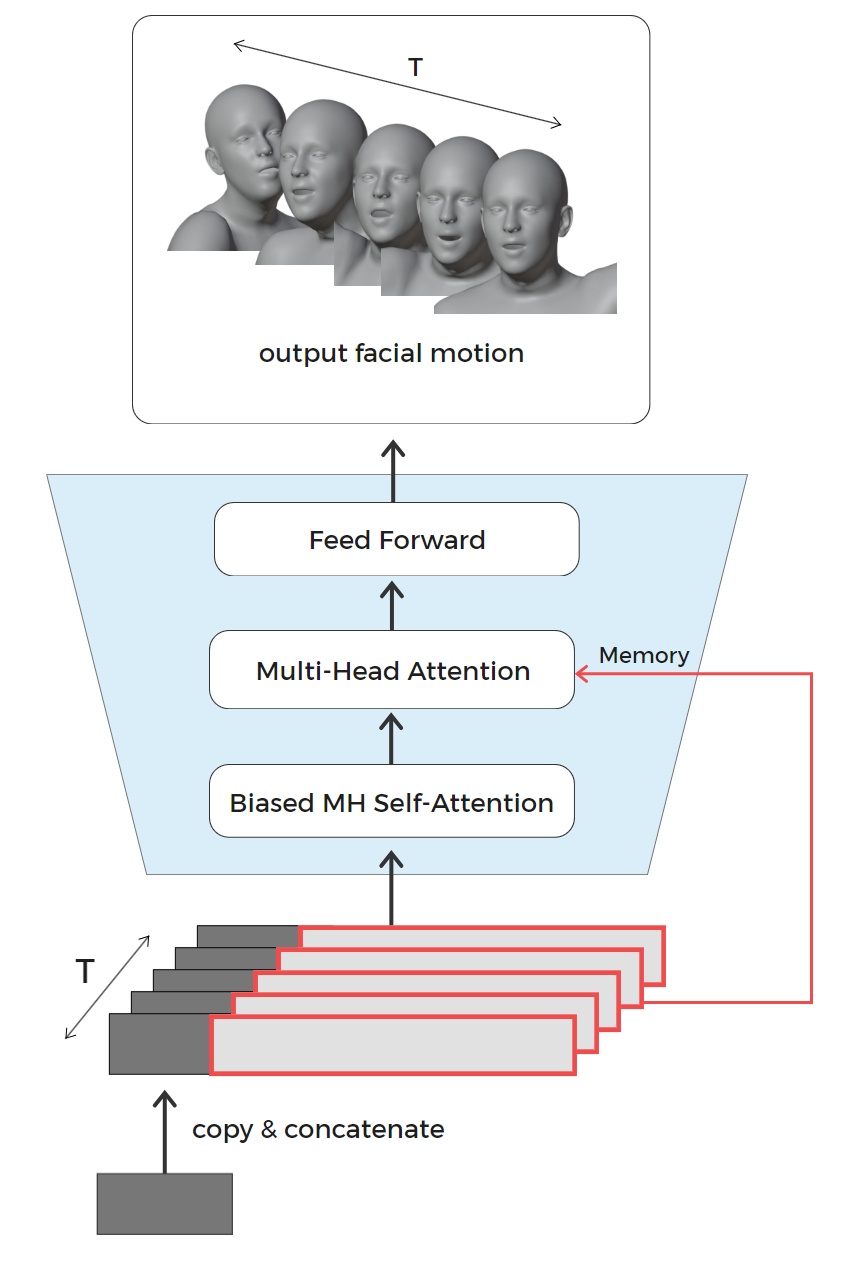}
    \caption{Facial Motion Generator architecture.}
    \label{fig:generator}
\end{figure}

\section{Dataset Details}

The Motion-X dataset \cite{Lin23} is a large-scale 3D whole-body motion dataset in SMPL-X format, comprising approximately 144 hours of motion data across over 81,000 motion clips. While some data in the Motion-X dataset combines body motion with synthesized facial motion, we excluded such data and used only a subset where facial and body motions were captured together. Specifically, we utilized about 57 hours of data from subsets such as HAA500, AIST, HuMMan, IDEA400, and Online Videos.

In some cases, the expression parameters within the FLAME parameters exhibited excessively large ranges, resulting in exaggerated facial expressions compared to those of real humans. To address this, we clipped the range of expression parameters to between -5 and 5 for training purposes.

Although the Motion-X dataset includes emotion labels such as neutral, happy, disgust, surprise, sad, angry, fear, and contempt, the B2F model is not dependent on emotion label information during training. Therefore, the labeled emotion information was not utilized in the training process.

\section{Loss Function Details}
{\color{rev}

The overall loss function, referenced as Equation 1 in the paper, is defined as follows:
loss function:
\begin{equation}
\mathcal{L} = \lambda_1 \mathcal{L}_{\text{recon}} + \lambda_2 \mathcal{L}_{\text{align}} +  \lambda_3 \mathcal{L}_{\text{KL}}+ \lambda_4 \mathcal{L}_{\text{consi}} + \lambda_5 \mathcal{L}_{\text{cross}}.
\label{eq:total-loss}
\end{equation}

The B2F model, used to generate all experimental results, was trained with the following weight configuration: \(\lambda_1 = 5.0\), \(\lambda_2 = 0.5\), \(\lambda_4 = 0.5\), and \(\lambda_5 = 0.1\). For the KL term, \(\lambda_3\) was not fixed but dynamically scheduled throughout training.

Specifically, \(\lambda_3\) followed a three-phase schedule: it linearly increased from 0 to a maximum value of 0.3 over the first 25\% of training epochs (warm-up), remained constant for the next 25\% (hold), and then linearly decayed back to 0 over the remaining 50\% (decay). This strategy prevents early KL dominance, allowing the model to learn stable and expressive latent representations without posterior collapse.
}

The $\mathcal{L}_{\text{recon}}$ referenced as Equation 2 in the paper is defined as follows:

\begin{equation}
\begin{aligned}
\mathcal{L}_{\text{recon}} 
= &\ \text{MSE}_\text{recon} \left( \mathbf F_\textrm A, \text{B2F}(\mathbf B_\textrm A, \mathbf F_\textrm A, \mathbf S_\textrm A) \right) \\
= &\ \text{MSE} \left( \mathbf F_\textrm A^{\text{exp}}, \text{B2F}(\mathbf B_\textrm A, \mathbf F_\textrm A, \mathbf S_\textrm A)^{\text{exp}} \right) + \\
&\ \lambda_{\text{jaw}} \cdot \text{MSE} \left( \mathbf F_\textrm A^{\text{jaw}}, \text{B2F}(\mathbf B_\textrm A, \mathbf F_\textrm A, \mathbf S_\textrm A)^{\text{jaw}} \right),
\end{aligned}
\end{equation}

where the superscripts \( \text{exp} \) and \( \text{jaw} \) refer to the expression and jaw components of the FLAME parameters, respectively. The weight \(\lambda_{\text{jaw}}\) is used to adjust for the scale difference between the expression and jaw parameters. The B2F model, used to produce all experimental results, was trained with \(\lambda_{\text{jaw}} = 1000\).

\section{FLAME-to-ARKit Converter Details}

\begin{figure}
\centering
\includegraphics[trim=0 120 0 150, clip,width=1.\linewidth]{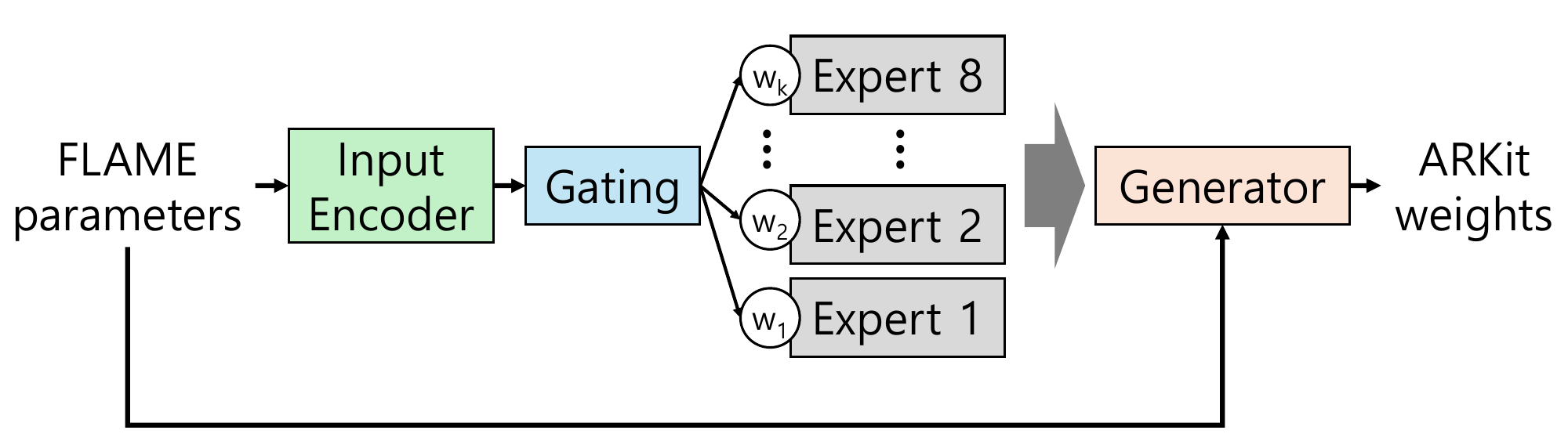}
\caption{FLAME-to-ARKit Converter architecture.
}
\label{fig:converter}
\end{figure}
Our FLAME-to-ARKit converter employs a parameter-blended Mixture of Experts (MoE) architecture, where the generator network produces the output by blending the parameters of expert networks based on the blending weights generated by a gating network (Figure~\ref{fig:converter}).

The input to our converter is the 103-dimensional FLAME parameters used in the training dataset, the BEAT dataset \cite{liu_beat_2022}, and the output is the 51-dimensional ARKit blendshape weights. The 103-dimensional FLAME input is first processed by an input encoder, which transforms it into a 32-dimensional input for the gating network. The gating network generates blending weights, which are used to blend the parameters of 8 expert networks, thereby determining the parameters of the generator network. The generator network then takes the 103-dimensional FLAME input and produces the 51-dimensional ARKit blendshape weights as output.
The input encoder is composed of a two-layer MLP with 64 and 32 hidden units. The gating network consists of a two-layer MLP with 128 and 128 hidden units. Both the expert networks and the generator network are implemented as three-layer MLPs with 128 hidden units in each layer. We observed that the parameter-blended MoE architecture significantly outperforms a simple architecture composed solely of MLP layers.

The training data consists of ARKit-based facial motion data from the BEAT dataset \cite{liu_beat_2022}, transformed into FLAME facial motion data using the ARKit-to-FLAME transformation matrix released in \cite{liu_emage_2024}\footnote{mat\_final.npy from https://drive.google.com/drive/folders/1ukbifhHc85qWTzspEgvAxCXwn9mK4ifr, accessed December 30, 2024}.
Our converter is trained using the MSE loss between the predicted FLAME parameters and the ground truth FLAME parameters. To emphasize mouth motion transformations, a weight of 500 was applied to the MouthClose blendshape value during MSE calculation.

Since the B2F model outputs the 53-dimensional FLAME parameters corresponding to the FLAME format of the Motion-X dataset \cite{Lin23}, during inference, only the first 50 dimensions of the 100-dimensional expression parameters from the 103-dimensional FLAME input are replaced with the 50-dimensional expression parameters output by the B2F model.

\section{Experimental Setup and Training Details}

All experiments and policy training were conducted on a desktop system equipped with an AMD Ryzen 7 7800X3D CPU and an NVIDIA GeForce RTX 4070 GPU, with the training taking approximately 14 hours. The batch size was set to 32, and the AdamW optimizer was used with an initial learning rate of $1.0 \times 10^{-4}$ and an initial weight decay of $5.0 \times 10^{-5}$.

{\color{rev}
\section{Addtional Experiment}

\paragraph*{Effect of Latent Code Perturbations on Style Expression}

\begin{figure*}
    \centering
    \includegraphics[trim=0 0 0 0, clip, width=1.\linewidth]{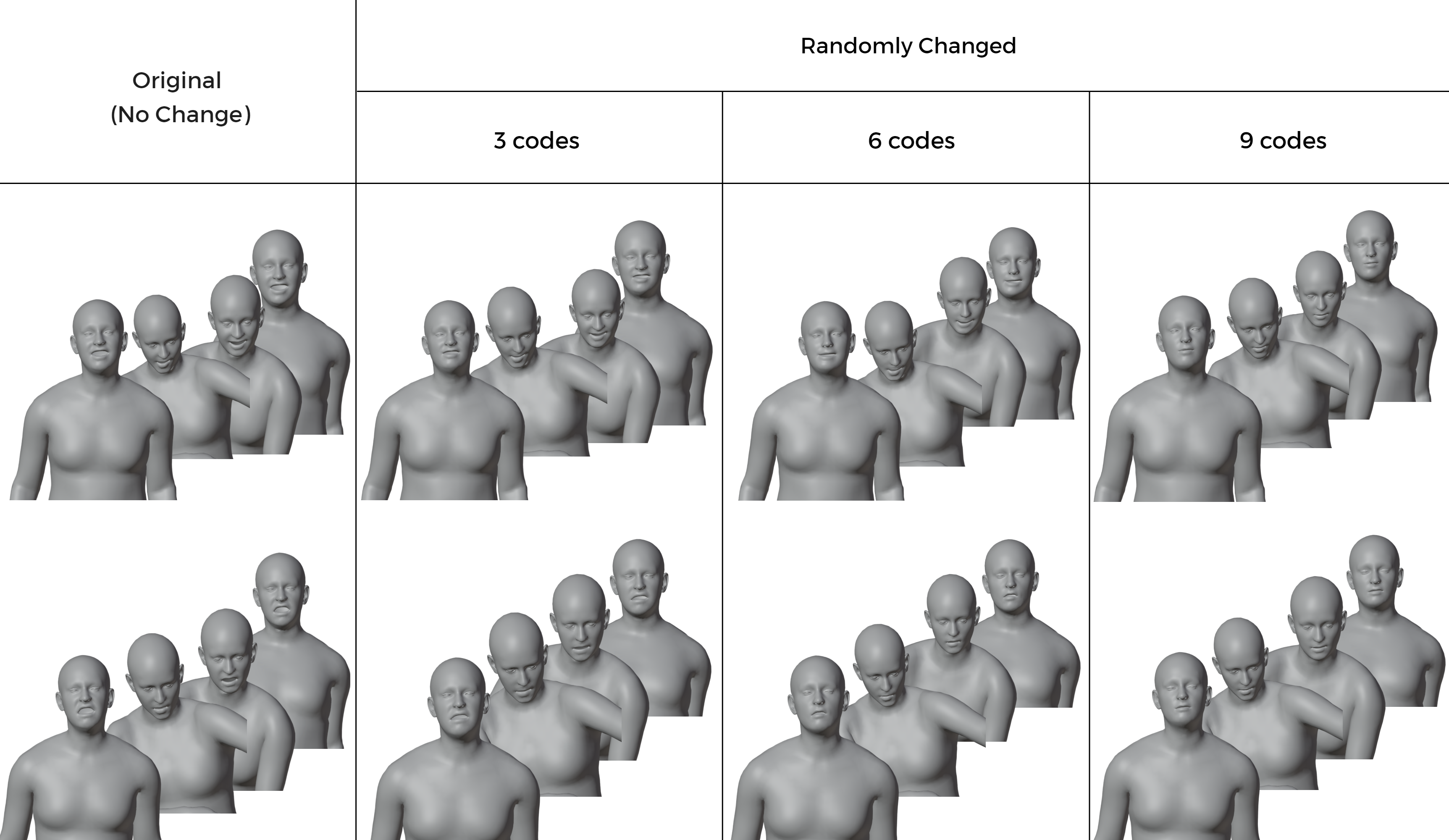}
    \caption{
Effect of latent code perturbations on style expression.
We randomly altered the active entries in a subset of the 12 one-hot vectors used in the discrete style code. As the number of modified vectors increases (3, 6, 9 out of 12), the resulting facial expressions show varying degrees of stylistic change. While not a quantitative evaluation, this visualization suggests that the style representation is responsive to local modifications in the latent code.
    }
    \label{fig:results-discrete}
\end{figure*}

To explore how the discrete latent code influences generated facial motion, we conducted a simple perturbation test.
The style encoder outputs $D=12$ one-hot vectors via hard sampling from categorical distributions, which are then flattened into the final style embedding.
In this test, we randomly selected a subset of these vectors and altered the index of the active (1-valued) entry in each chosen vector.

As shown in Figure~\ref{fig:results-discrete}, modifying more vectors led to more noticeable changes in the resulting expressions.
While this result is not intended as a rigorous evaluation, it provides a visual indication that localized edits in the latent code can affect the style of the generated motion.

}

\end{document}